%% file: FLchain_Survey_Manuscript.tex
\documentclass[journal,10pt,twocolumn,twoside]{IEEEtran}
\usepackage{mathtools}
\usepackage{empheq}
\usepackage{algpseudocode, algorithm}
\usepackage{algorithmicx}
\usepackage{subfig}
\usepackage[table]{xcolor}%
\usepackage{varwidth}
\usepackage{url}
\usepackage{multirow}
\usepackage{amssymb}
\usepackage{mathtools}
\usepackage{changes}
\usepackage[short]{optidef}
\usepackage{caption}
\usepackage{amsmath}
\usepackage{textcomp}
\usepackage{gensymb}
\usepackage{cite}
\usepackage{booktabs}
\usepackage{hyperref}
\usepackage{balance}
\usepackage{booktabs}
\usepackage{array}
\usepackage{balance}
\newcolumntype{P}[1]{>{\centering\arraybackslash}p{#1}}

\begin{document}

\title{Federated Learning Meets Blockchain in Edge Computing: 
Opportunities and Challenges }

\author{Dinh C. Nguyen,	Ming Ding, Quoc-Viet Pham, Pubudu N. Pathirana, Long Bao Le, \\ Aruna Seneviratne, Jun Li, Dusit Niyato,~\IEEEmembership{Fellow,~IEEE}, and H. Vincent Poor,~\IEEEmembership{Life Fellow,~IEEE}
	
	\thanks {*This work was supported in part by the CSIRO Data61, Australia.}
	\thanks{Dinh C. Nguyen and Pubudu N. Pathirana are with the School of Engineering, Deakin University, Waurn Ponds, VIC 3216, Australia (e-mails: \{cdnguyen, pubudu.pathirana\}@deakin.edu.au).}
	\thanks{Ming Ding is with the Data61, CSIRO, Australia (email: ming.ding@data61.csiro.au).}
	\thanks{Quoc-Viet Pham is with the Korean Southeast Center for the 4th Industrial Revolution Leader Education, Pusan National University, Busan 46241, Korea (e-mail: vietpq@pusan.ac.kr).}
	\thanks{Long Bao Le is with the Institut National de la Recherche Scientifique, University of Quebec, Montreal, QC H5A 1K6, Canada (email: long.le@emt.inrs.ca).}
	\thanks{Aruna Seneviratne is with the School of Electrical Engineering and Telecommunications, University of New South Wales (UNSW), NSW, Australia (email: a.seneviratne@unsw.edu.au).}
	\thanks{Jun Li is with the School of Electrical and Optical Engineering, Nanjing University of Science and Technology, Nanjing 210094, China (e-mail: jun.li @njust.edu.cn).}
	\thanks{Dusit Niyato is with the School of Computer Science and Engineering, Nanyang Technological University, Singapore (email: dniyato@ntu.edu.sg).}
	\thanks{H. Vincent Poor is with the Department of Electrical Engineering, Princeton University, Princeton, NJ 08544 USA (e-mail: poor@princeton.edu).}
}

\IEEEtitleabstractindextext{
\begin{abstract}
Mobile edge computing (MEC) has been envisioned as a promising paradigm to handle the massive volume of data generated from ubiquitous mobile devices for enabling intelligent services with the help of artificial intelligence (AI). Traditionally, AI techniques often require centralized data collection and training in a single entity, e.g., an MEC server, which is now becoming a weak point due to data privacy concerns and high data communication overheads. In this context, federated learning (FL) has been proposed to provide collaborative data training solutions, by coordinating multiple mobile devices to train a shared AI model without exposing their data, which enjoys considerable privacy enhancement. To improve the security and scalability of FL implementation, blockchain as a ledger technology is attractive for realizing decentralized FL training without the need for any central server. Particularly, the integration of FL and blockchain leads to a new paradigm, called \emph{FLchain}, which potentially transforms intelligent MEC networks into decentralized, secure, and privacy-enhancing systems. This article presents an overview of the fundamental concepts and explores the opportunities of FLchain in MEC networks. We identify several  main topics in FLchain design, including communication cost, resource allocation, incentive mechanism, security and privacy protection. The key solutions for FLchain design are provided, and the lessons learned as well as the outlooks are also discussed. Then, we investigate the applications of FLchain in popular MEC domains, such as edge data sharing, edge content caching and edge crowdsensing. Finally, important research challenges and future directions are also highlighted.  
\end{abstract}

}

\maketitle
\IEEEdisplaynontitleabstractindextext
\IEEEpeerreviewmaketitle

\markboth{Accepted at IEEE Internet of Things Journal}%
{}

\input{Introduction.tex}

\input{State-of-Art.tex}

\input{BlockchainFL_Design.tex}
\input{BlockchainFL_Applications.tex}
\input{Challenges_Future-Directions.tex}

\input{Conclusion.tex}
\balance
\bibliography{Ref}
\bibliographystyle{IEEEtran}

\end{document}

%% file: Introduction.tex
\section{Introduction}
\label{Sec:Introduction}

The rapid development of the Internet of Things (IoT) has transformed business and customer services in many aspects of modern life with ubiquitous sensing and computing capabilities of mobile devices such as smartphones, laptops, and tablets \cite{1}. Aiming at obtaining the insights of data generated from these mobile devices, artificial intelligent (AI) disciplines such as machine learning (ML) have been widely exploited to train data models for enabling intelligent applications \cite{9200330}.
{To accommodate the unpredictable demands of computation and storage for AI services, mobile edge computing (MEC) has been proposed for data training with low latency, by exploiting resources at the network edge near the data sources \cite{2}.}
Given the increasing volume of IoT data and the growing concern of data privacy in next-generation wireless networks, implementing centralized AI training at the central server may be no longer suitable. Federated learning (FL) has recently emerged as a 
{distributed} 
AI approach, by coordinating multiple devices to perform AI training without sharing raw data for {privacy enhancement} and network resource (e.g., bandwidth) savings. For example, in the context of MEC networks, distributed mobile devices can collaborate with an MEC server to perform an FL process \cite{3}, by training data locally and exchanging learning parameters with the {aggregation server} in a number of communication rounds until the global training is complete. {In this way, users do not need to offload their raw data to the server, which is able to enhance user privacy and save communication resources required for data transmission.}

However, some issues still remain in current FL systems. Firstly, users need to fully trust the MEC server for model aggregation, but this is not always achieved in realistic wireless networks. Secondly, although FL can help enhance user privacy, the transmission of learning parameters to the MEC server is vulnerable to security bottlenecks such as malicious threats that can modify or steal the information of local updates \cite{4}. Moreover, the reliance of an MEC server for model aggregation introduces single-point-of-failure bottlenecks once the server is attacked, which consequently disrupts the entire FL system. Thirdly, given the high scalability of modern edge computing networks, a single MEC server cannot manage to aggregate all updates offloaded from millions of devices. Therefore, there is an urgent need to develop a more decentralized FL approach without using a central server so as to solve security and scalability issues for enabling the next generation intelligent edge networks.

Fortunately, blockchain as a ledger technology \cite{507} can provide attractive solutions for FL-based intelligent edge computing \textcolor{black}{due to its unique features such as decentralization, immutability, and traceability}. Indeed, a blockchain is a form of blocks linked by hash values under the control of a consensus mechanism, e.g., Proof-of-Work (PoW), enabled by miners which mine blocks with digital signature to make linked blocks immutable against modifications and alterations \cite{507}. \textcolor{black}{By using blockchain, FL can be implemented via decentralized data ledgers without requiring any central server which mitigates risks of single-point failures, and any update events and user behaviours are traced by all network entities in a transparent manner. Furthermore, one can easily trace the origin where a model parameter is modified or updated during the training process through transaction logs, which cannot be satisfied in traditional FL systems.} Particularly, the integration of FL and blockchain creates a new paradigm called \emph{FLchain}, which potentially
transforms intelligent edge networks with decentralized and secure natures \cite{6}, \cite{7}. In FLchain, each device can act as a client with equal rights to update and aggregate the learning model in a decentralized manner. More specifically, each client first initializes a data model and computes learning parameters. Then, it uploads the computed update to a group of miners in a form of transactions. Here, miners combine transactions (which contain local updates) created by clients into a block after a period of time, which is then verified by miners through a mining process. Once the block is mined, it is appended to the blockchain and broadcast to the entire network. Each client downloads the block, which contains the updates of other clients, and computes a new version of the global model. The process iterates until the global loss function converges or the desired accuracy is achieved. In this way, blockchain helps eliminate the need for a central server which thus potentially alleviates communication costs and achieves better intelligent network scalability. Moreover, blockchain provides high security for FL training through immutable block ledgers. 

\subsection{Comparison and Our Contributions}
Some research efforts have been devoted to FL and blockchain in edge computing. The works in \cite{9}, \cite{10}, \cite{11} discussed the key concepts of FL and its protocol, architectures, and applications in wireless networks. The survey in \cite{12} presented the use of FL in mobile edge networks, focusing on challenges of FL implementation in edge computing and possible applications of FL in edge network optimization. Another study in \cite{13} provided an overview of FL in fog computing
along with a discussion of fundamental FL theories in learning and model compression. Moreover, solutions and methods for data preservation in FL implementation were explored in \cite{14}. Meanwhile, the potential of blockchain in edge computing was investigated in \cite{15} through a holistic survey on related issues such as scalability, resource management and network security. The use of blockchain in edge computing was also analyzed in \cite{16} from the perspective of consensus and scalability.

Although FL and blockchain in edge computing have been extensively studied in the literature, there is no any existing work to survey the integration of FL and blockchain in edge computing, to the best of our knowledge. To fill this research gap, in this article we present an extensive survey on the integrated FL-blockchain for enabling intelligent and secure edge networks, 
{beginning with an introduction to FL, blockchain, and a generic FLchain architecture.}
We then focus on analyzing important design issues and use cases of FLchain in edge computing, including communication cost, resource allocation, incentive learning, security and privacy protection. Particularly, we provide an extensive survey on the use of FLchain in various applications in edge computing, such as edge data sharing, edge content caching, and edge crowdsensing. Possible research challenges and future directions are also outlined. To this end, the key contributions of this article are highlighted as follows:
\begin{enumerate}
	\item 	We provide an overview of the fundamentals of FL and blockchain and {propose a novel FLchain architecture applicable to edge computing networks.}
	\item 	{We identify and discuss the technical issues in FLchain such as} communication cost, resource allocation, incentive learning, security and privacy protection.
	\item 	We analyze the opportunities brought by FLchain in edge computing {for several application domains}, such as edge data sharing, edge content caching, and edge crowdsensing. 
	\item 	Finally, we outline key research challenges and discuss possible future directions toward the full realization of FLchain in edge computing. 
\end{enumerate}

\subsection{Structure of The Survey}
The remainder of the article is organized as follows. Section~\ref{Sec:State-of-Art} presents the state-of-the-art and fundamentals of FL and blockchain. In Section~\ref{Sec:Threat}, we describe the threat models and then explain the motivations behind the integration of FL and blockchain. A generic FLchain architecture is also proposed where the network components and the working concept are presented. The design and some key use cases of FLchain implementation in edge computing are discussed in Section~\ref {Sec:BlockchainFL_Design}, while the applications of FLchain in edge computing are analyzed in Section~\ref{Sec:BlockchainFL_Applications}. The key research challenges and future directions are highlighted in Section~\ref{Sec:Challenges_Future-Directions}. Finally, Section~\ref{Sec:Conclusion} concludes the article.

%% file: State-of-Art.tex
\section{{Federated Learning and Blockchain: State-of-the-Art}}
\label{Sec:State-of-Art}
In this section, we present the fundamentals of blockchain and FL and their recent developments toward the FLchain integration.

\subsection{Federated Learning}
Since the invention in 2016 \cite{17}, FL has transformed many applications by providing distributed AI solutions at the network edge with the cooperation of multiple mobile devices. Instead of offloading all raw data to a cloud or a data center to perform AI training, FL enables distributed learning by allocating AI functions, e.g., AI model training, directly to the local devices to build a shared global model at an aggregator, e.g., an MEC server at a base station (BS) or an access point (AP) \cite{18}. Following the recent advances in mobile hardware and increasing concerns of user privacy, FL is particularly attractive for various intelligent edge services such as smart transportation and smart healthcare \cite{19}. For example, in smart vehicular networks, vehicles can act as learning clients to train local models and collaborate with a road side unit (RSU) to build traffic prediction models, aiming to build a comprehensive vehicle routing map for reducing traffic congestion \cite{20}. As a BS cannot collect all data from distributed mobile devices for AI/ML training, FL is of paramount importance in realizing full intelligence in the next generation of mobile edge networks. FL allows devices and the BS to train a common model while raw datasets are kept locally at users where the data reside. 

In an FL process, we denote the set of mobile devices as $\mathcal{K}$. Each device $k \in \mathcal{K}$ participates in training a shared AI model by using its own dataset $D_{k \in \mathcal{K}}$. More specifically, in every communication round, each device trains a local model and calculates an update $\textbf{w}_k$  by minimizing a loss function {$\mathcal{F}(\textbf{w}_k)$}  
as follows:
\begin{equation}
\textbf{w}_k^* = \arg\min \mathcal{F}( \textbf{w}_k), k \in \mathcal{K}. 
\end{equation}
Here, the loss function of different FL algorithms can be different \cite{21}. For example, given a set of input-output pairs $\{x_i,y_i\}_{i=1}^K$, the loss function $\mathcal{F}$ of a linear regression FL model can be determined as $\mathcal{F}( \textbf{w}_k) = \frac{1}{2}\left(x_i^T \textbf{w}_k -y_i\right)^2$. Then, each device $k$ uploads its computed update $\textbf{w}_k$ to the MEC server, which then aggregates and calculates a new version of global model as
\begin{equation}
\textbf{w}_G = \frac{1}{\sum_{k \in \mathcal{K}}|D_k|}\sum_{k=1}^{K}|D_k|\textbf{w}_k.
\end{equation}
This global model is then downloaded to all devices for the next round of training until the global learning is complete. 

It is noticed that in the classical FL, the global model $\textbf{w}_G$ is computed and updated by the MEC server. However, in FLchain, the global model computation is performed directly at devices in a decentralized manner via blockchain. In the next section, we will analyze system designs and use cases of FLchain in various applied domains. 

\subsection{Blockchain}
\subsubsection{Fundamentals}
Blockchain is basically a public, trusted and shared ledger running on a peer-to-peer (P2P) network. The key idea behind the blockchain concept is its decentralization, that is, data on blockchain is not controlled by any single entity \cite{22}. Instead, all blockchain nodes such as mobile devices and MEC servers in edge networks, have the equal right to verify and manage the data stored in blockchain enabled by consensus mechanisms. This decentralized feature makes blockchain resistant to data modifications or attacks. Moreover, the elimination of central server avoids the risk of single-point failures, thus improving the reliability and stability of blockchain systems.  
\begin{figure}
	\centering
	\includegraphics[width=0.95\linewidth]{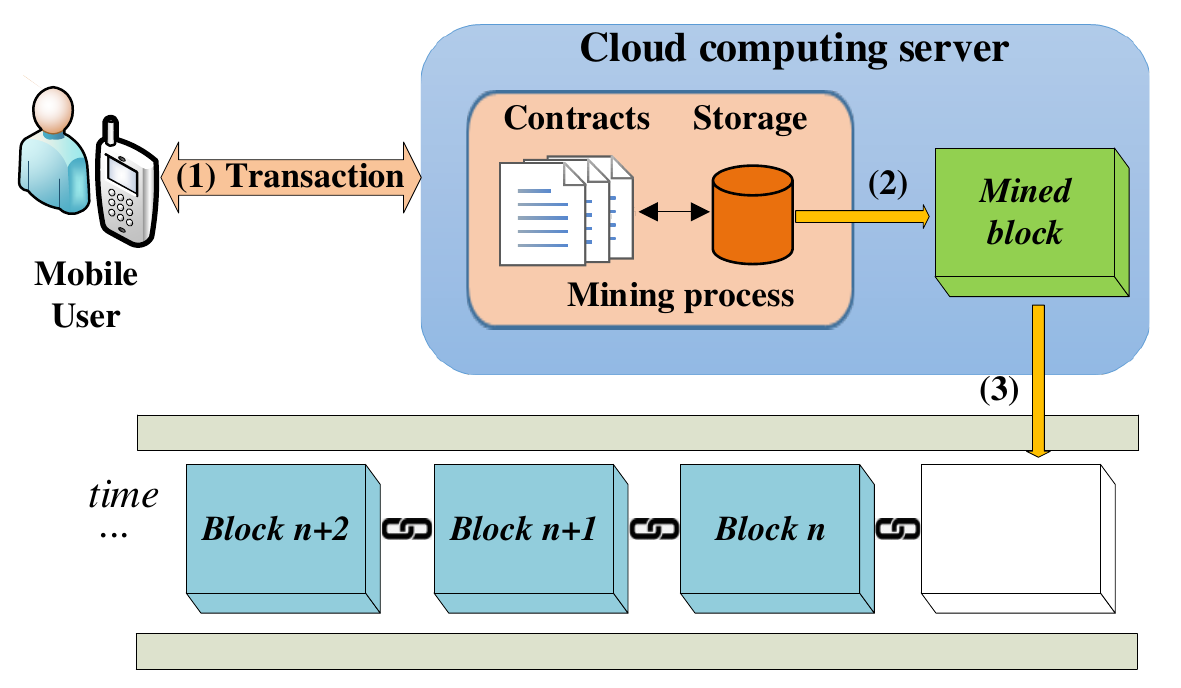}
	\caption{{\color{black}The transaction process on blockchain with cloud computing: (1) A user sends a transaction to smart contracts to request data, (2) A new block is created to represent the verified transaction, (3) A mined block is appended to the blockchain.}}
	\label{Blockchain_operation}
	\vspace{-0.1in}
\end{figure}
{\color{black}A typical blockchain operation is described in Fig.~\ref{Blockchain_operation} and explained as follows.
\begin{itemize}
	\item \textit{Step 1:} The mobile user uses its wallet account, which includes a private key and a public key, to create a transaction with metadata (i.e. user ID), user signature and timestamp. Due to the resource constraints of mobile devices, extensive-computational tasks including mining, smart contract execution can be organized and implemented on cloud computing. The user then submits a transaction to the cloud computing for a certain request, such as IoT data task processing, user authentication, cloud service queries. 
	\item \textit{Step 2:} The cloud releases available resources and process the request from the user. Smart contracts automatically perform transaction authentication, user verification, and trading (i.e. asset trading among users and service providers). Besides, cloud mining empowered blockchain miners (i.e. virtual machine) verifies the user transaction. This transaction is added to the pool of unconfirmed transactions where it waits for its turn to be picked up by a miner. A network of virtual miners can achieve on an agreement via a consensus mechanism such as PoW. 
	\item \textit{Step 3:} The fastest miner which verifies the data block will send the signature to other miners for validation. If all miners achieve an agreement, the validated block with its signature is then appended to the blockchain in a chronological order. Finally, all network entities receive this block and synchronize the copy of the blockchain. 
\end{itemize}
The operations of blockchain also incur costs in terms of latency and energy consumption \cite{22}. In fact, the execution of blockchain mining tasks, e.g., block verification and information exchange among miners requires large energy sources to be consumed. To be clear, in order to append a new transaction to the blockchain, a blockchain user or a miner needs to run a mining puzzle, e.g., PoW which is generally complicated and requires vast computing and storage resources. This issue becomes more important in the mobile blockchain network where mobile devices often have limited battery and memory which make them challenging to accommodate all mining tasks. Moreover, the repeated information exchange among miners in the mining group also requires large bandwidth resources and results in high communication latencies which potentially degrades the overall performance of FLchain training, e.g., delayed model aggregation. Therefore, it is highly important to carefully consider latency and energy aspects to minimize possible costs caused by the blockchain adoption in FLchain systems. 
\subsubsection{Blockchain Categories}
Blockchain can be categorized into three main types, including public (or permission-less), private (or permissioned), and consortium blockchain, which are explained as follows: 
\begin{itemize}
	\item \textit{Public Blockchain}: is an open network which allows everyone to join and make transactions as well as participate in the consensus process. The best-known public blockchain consists of Bitcoin and Ethereum with open source and smart contract blockchain platforms. For example, Ethereum \cite{15}  is a distributed public blockchain network similar to most platforms such as Bitcoin. A big advantage of Ethereum is its capability of automatic digital asset management through smart contracts which run on Ethereum virtual machine. Blocks are verified and appended to the blockchain by a PoW algorithm managed by miners to achieve a secure, tamper-resistant consensus among all nodes in the network. Each block needs a certain amount of gas, a currency in Ethereum, to be consumed as part of its execution, and pays for miners as the reward of mining the block. 
	
	\item \textit{Private Blockchain}: is an invitation-only network managed by an authority and all activities such as transaction writing and retrieval in the blockchain need to be permissioned by a validation mechanism. While the private blockchain approach assumes that the network is operated by a single entity, consortium blockchains operates under the management of a group of owners. They restrict user access to the network and the transactions performed by the network members. Due to the ability to provide collaborative control in the network, consortium blockchains enable interconnected business transformation among organizations and innovative business models. 
	
	\item	\textit{Consortium Blockchain}: is a blockchain platform governed by multiple organizations instead of only a single organization. A popular consortium blockchain platform is Hyperledger Fabric \cite{507}. Hyperledger Fabric represents as a consortium blockchain platform that was founded by the Linux Foundation in 2015. Similar to other blockchain technologies, Hyperledger Fabric has a ledger, uses smart contracts called chaincode, and coordinates participants in transaction organization. However, the concept of managing network access is different. Rather than an open permission-less system that enables the participation of all members in the network, the members of a Hyperledger Fabric network must be trusted by a membership service provider. Fabric takes a novel architectural approach and revamps the way blockchains cope with nondeterminism, resource exhaustion, and performance attacks. Fabric can also create channels, which enable a group of participants to establish a separate ledger of transactions. To achieve consensus in the network, Fabric uses Practical Byzantine Fault Tolerance (PBFT) which is well suitable for enterprise consortiums where members are partially trusted.
\end{itemize}}

\section{Threat Models, Integration Motivations and Proposed FLchain Architecture}
\label{Sec:Threat}
In this section, we describe the threat models, and motivations of the integration of FL and blockchain. Then, we introduce and analyze the proposed the FLchain architecture. 
{\color{black}\subsection{Threat Models and Security Requirements}
\subsubsection{Threat Models}
We consider two potential threat types in FLchain design as follows.

\begin{itemize}
	\item \textit{Insider Threats:} In the FLchain system, the MEC servers can be semi-trusted in the training process.   Under this assumption, the MEC servers may be honest but curious about the parameter updates and thus can infer some sensitive information from transactions on the blockchain. More specifically, although the data is not explicitly shared in the original format, it is still possible for curious MEC servers to steal the training data from gradients and reconstruct the raw data approximately, especially when the architecture and parameters are not completely protected \cite{4}. Moreover, malevolent clients can exploit and learn data structure such as image pixels extracted from the global model update without the consent of other clients and MEC servers. 
	
	\item \textit{External Threats:} At the client side, an adversary can modify data feature or inject an incorrect subset of data in the original dataset to embed backdoors into the model, aiming to adjust the training objective of local clients. An attacker can also compromise some client devices, and the attacker manipulates the local model parameters on the compromised client devices during the learning process which results in errors in the global model update \cite{508}.  Moreover, adversaries can deploy attack on the wireless communication channels during the FL training to gain personal information of clients. For example, an adversary can retrieve the sensitive user information mixed in the parameter update package, e.g., age and user preference \cite{505}. Further, external eavesdroppers can gain unauthorized access to the MEC servers in order to take control of the model update aggregation procedure. 
\end{itemize}
\subsubsection{Security Requirements}
Firstly, it is important to provide a high degree of privacy protection for the FL training, aiming to ensure that data information is safe and encourage users to join the data training. This can be done by using perturbation techniques \cite{506} such as differential privacy can be used to protect training datasets against data breach, by constructing composition theorems with complex mathematical solutions. Secondly, threat management and attack defense solutions are desired to solve security issues from both the client and MEC server sides. For example, it is required to deploy attack detection mechanism at the aggregation server to evaluate the weight contribution of each client, aiming to filter out adversarial clients and detect attacks such as model update poisoning attacks, data poisoning attacks, and evasion attacks in each communication round. Thirdly, it is highly needed for ensuring security and privacy on the wireless communication channels during the parameter exchange and update broadcast. Data encryption, communication authentication and secure ledger configurations using blockchain and smart contracts \cite{511} can be useful solutions to meet these security requirements in the FL communications. Finally, mobile devices in the FLchain system are able to establish transactions and communicate with their associated MEC servers for secure FLchain training. Each device should also devote resources (e.g., storage and computation) to train learning models and run mining for extra profits. }

\subsection{Integration Motivations}
With its unique properties, blockchain has the great potential to improve security for FL in edge networks. \textcolor{black}{Specifically, the use of decentralized blockchain allows to eliminate the need for a central server in FL training \cite{25}. Instead, a shared immutable ledger is used to aggregate the global model and distribute global updates to learning clients for direct computation at devices. The decentralization of model aggregation not only mitigates the risk of single-point failures for better training reliability but also reduces the burden posed on the central server in global model aggregation, especially when edge networks have numerous clients.} The learning updates are appended to immutable blocks for information exchange among clients during the training which ensures high security for training against external attacks. The replication of blocks to the entire network also allows all clients to verify and trace the training progress so as to ensure high trust and transparency for the FLchain system. Moreover, the elimination of a central server for FLchain model aggregation potentially alleviates communication costs and attract more mobile users to participate in data training based on its decentralized network topology, which in return enhances the scalability of mobile edge networks. 

\textcolor{black}{Motivated by these unique benefits, recently many network architectures have been proposed by integrating blockchain in FL.} For example, an integrated FL-blockchain scheme is proposed in \cite{25}, where both local training and global model updating are performed at mobile devices via the blockchain in the MEC network. Another integrated FL-blockchain architecture is introduced in \cite{27} which allows FL clients (i.e., vehicles) to collaboratively train AI models and exchange aggregated updates through blockchain miners. FL is also integrated with blockchain for an MEC-based IoT network \cite{29}, aiming to coordinate IoT users to train neural networks and share training parameters with MEC servers at BSs for cooperatively solving a data relaying optimization problem. A similar integrated FL-blockchain architecture is considered in \cite{53} to collaborate IoT devices and MEC servers at BSs. In particular, a shared global model is learned using distributed local datasets under the control of a blockchain which also provides model verification in a decentralized manner.
\subsection{Proposed FLchain Architecture}
\begin{figure}
	\centering
	\includegraphics[width=0.99\linewidth]{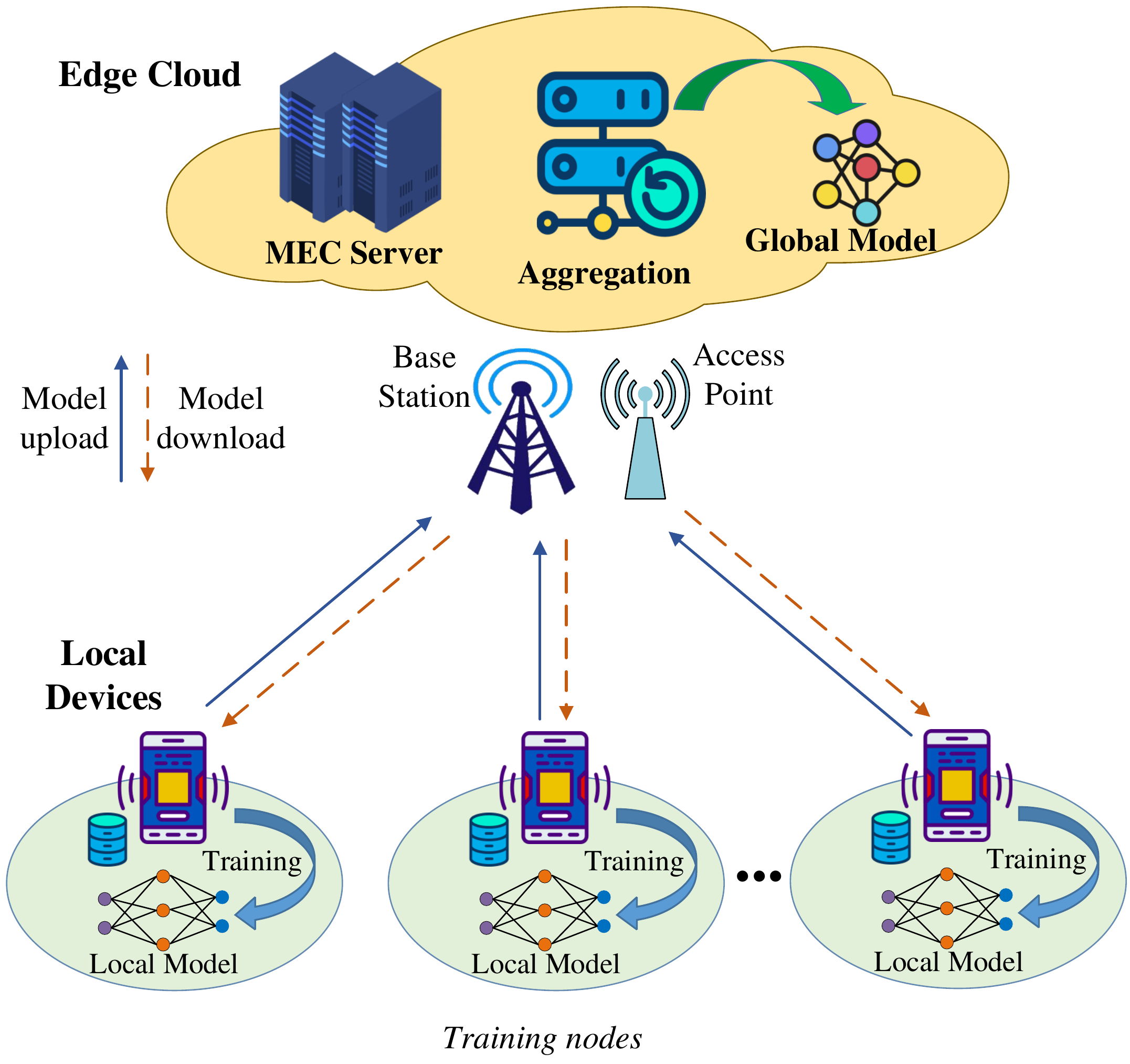}
	\caption{\textcolor{black}{The classical FL architecture.}}
	\label{Fig:FL_Classic}
	\vspace{-0.1in}
\end{figure}

\begin{figure*}
	\centering
	\includegraphics[width=0.97\linewidth]{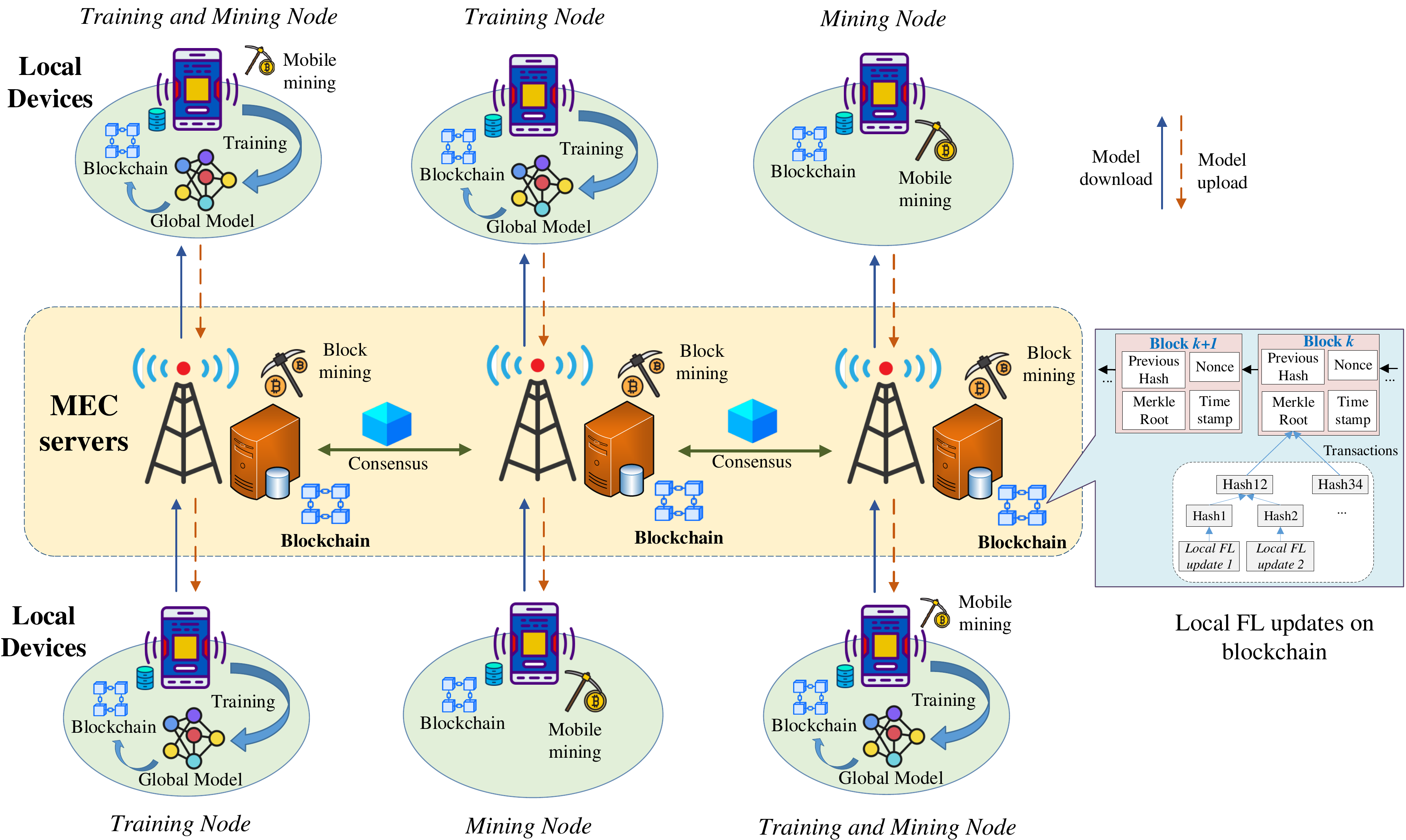}
	\caption{The proposed FLchain architecture. The MEC servers are responsible for blockchain mining, while mobile devices can participate in local training or mining, or implement both training and mining tasks. {\textcolor{black}{Through the mining process, the local updates of all mobile devices are aggregated and stored in a block for sharing across the network in each communication round, and then devices download the generated block to compute the global model locally.}}}
	\label{Fig:FL_Concept}
	\vspace{-0.1in}
\end{figure*}

\textcolor{black}{Inspired by the existing integrated FL-blockchain schemes, this paper introduces a more generic FL-blockchain architecture or FLchain,} which covers the most important features of current FL-blockchain system designs. First, we recall the key concept of the classical FL which has been widely used in distributed AI systems. In general, there are two key entities in a classical FL system, including distributed devices and an aggregator,
as illustrated in Fig.~\ref{Fig:FL_Classic}. Conceptually, the operational process of a classical FL system includes four steps:

\begin{enumerate}
	\item The MEC server at the edge cloud first selects a computing task, e.g., vehicular traffic analytics, along with task requirements, e.g., task classification or task prediction, and learning parameters, e.g., learning rate. Then, the MEC server chooses a subset of devices as learning clients for collaborative training. In practice, the MEC server can select different subsets of devices over different update rounds by using a suitable client scheduling mechanism to improve its training quality \cite{27}. 
	\item	The server initializes a training model and sends it to all clients to create a new round of training. Here, each client trains the model using its own dataset and calculates a new update.
	\item	Each client uploads its computed update to the MEC server for global computation. By aggregating all updates from clients, the server constructs a new global model in a fashion that minimizes a loss function \cite{8}. 
	\item	The server then broadcasts the computed global model to all clients for the next round of training. The FL process is iterated until the global loss function converges or the desired accuracy is achieved.
\end{enumerate}
From the learning procedure, we can see that the classical FL relies on a centralized server, i.e., an MEC server, for model aggregation, which suffers from malfunction and scalability issues in ubiquitous wireless networks. Moreover, such a centralized FL architecture cannot attract devices located far from the server for training which thus limits the learning performance of the overall system. 

In this context, FLchain comes as a strong alternative to decentralize the classical FL process by allocating global model computation directly to each device without the need for any centralized server. Particularly, FLchain provides unique security features for the FL training offered by immutable blockchain, which thus enables reliable intelligent decentralized edge networks. The proposed FLchain architecture is shown in Fig.~\ref{Fig:FL_Concept}, including a group of MEC servers and distributed devices in a connected blockchain network. {Due to the high computational capability, MEC servers are often selected to perform mining to maintain the blockchain network, while mobile devices can participate in local training or mining, or implement both training and mining tasks \cite{25}.} {The  general FLchain procedure} is summarized via the following steps.
\begin{enumerate}
	\item 	A group of MEC servers is initialized with their associated devices for a specific learning task (e.g., federated healthcare analytics) in edge networks. Each MEC server as a learning client devotes its resource for running blockchain consensus (or mining), and mobile devices join to run training algorithms in the FL process.
	\item	{Each training node computes a local model by using its own data and then transmits the local model to its associated MEC server via blockchain by creating a transaction.}
	\item	\textcolor{black}{The MEC servers collect transactions from their clients and store them according to a defined data structure (e.g., a Merkle tree \cite{nguyen2019integration}), and then create a block in a certain time slot. Each block is identified by a unique hash value along with a time stamp and a nonce which aims to prevent unauthorized reproductions of the block.} Then, the MEC servers participate in the mining process (e.g., PoW) to verify the newly created block and achieves a consensus among all the MEC servers. An MEC server can be chosen to act as a manager to coordinate the mining process in its time slot. {Note that devices also participate in mining blocks as full or partial nodes for extra profits.}
	\item	When the mining is done, the verified block is added to the blockchain and broadcast to all local devices via server-device communications. Now, the local model updates are stored securely in the blockchain.
	\item	Local devices download the block that contains all local updates of other devices. That allows each device to compute the global model directly at the local device {based on a pre-defined model aggregation rule such as weighted sum rule or error-based aggregation rule \cite{add1}}. In other words, the global model is computed locally instead of in the central server like in the traditional FL architecture. The training process is iterated until the global loss function converges or the desired accuracy is achieved.
\end{enumerate}

\begin{table*}
	\centering
	\caption{Comparison between the classical FL and FLchain. 	 }
	\label{Table:Comparison}
	{\color{black}
		\begin{tabular}{|P{1.2cm}|P{5.5cm}|P{4.6cm}|P{5cm}|}
			\hline
			{\textbf{FL Types}}& 	
			{\textbf{Benefits}}& 
			{\textbf{Drawbacks}}&
			{\textbf{Operational conditions}}
			\\
			\hline
			\multirow{7}{1.5cm}{Classical FL} &
			\vspace*{-1mm}
			\begin{itemize}
				\item 	Ability to train AI models at local devices without sharing raw data.
				\item	Ability to {enhance} user data privacy 
				\item	Ability to save network resources, e.g., bandwidth, transmit power. 
			\end{itemize}& 
			\vspace*{-1mm}
			\begin{itemize}
				\item 	Remain security issues (e.g., data attacks, single-point failures)
				\item	High communication delays for remote devices with the server. 
				\item 	{Non-transparent model aggregation and lack of incentive mechanisms.} 
			\end{itemize}& 
			\vspace*{-1mm}
			\begin{itemize}
				\item All devices and the server need to establish robust communications for collaborative training.
				\item Learning clients or devices need to trust the central server.
			\end{itemize}
			
			\\ \cline{2-4}
			\hline
			\multirow{6}{1.5cm}{FLchain} & 
			\vspace*{-1mm}
			\begin{itemize}
				\item 	Do not need a central server. 
				\item 	Eliminate single-point-of-failure risks.
				\item 	Can build trust among devices and servers
				\item	Improve the scalability of intelligent edge networks. 
			\end{itemize}& 
			\vspace*{-1mm}
			\begin{itemize}
				\item 	Possible latency and energy cost required by blockchain mining. 
				\item 	{Possible privacy risks from open update sharing.}
				\item Possible conflict between training data organization and blockchain storage in FLchain training.
			\end{itemize}& 
			\vspace*{-1mm}
			\begin{itemize}
				\item 	Mobile devices are able to train learning models and run mining for extra profits (as shown in Fig.~\ref{Fig:FL_Concept})).
				\item	A blockchain platform used in FLchain is able to establish a decentralized data network among by all mobile devices for shared training.  
			\end{itemize}
			\\ \cline{2-4}
			\hline
	\end{tabular}}
\end{table*}

{\color{black}In the FLchain training, the training examples may not be fully deterministic. For example, in federated online image classification tasks, clients may have different image datasets with varying pixel resolution which can be also updated continuously through sensing environments. Meanwhile, the transaction storage in blockchain requires exact content match where the hash value of a transaction must be unique to its content. Given that fact, how to achieve synchronization among training data organization and blockchain storage during the FLchain training is an important issue. A possible solution is to develop adaptive mechanisms for hash value generation with respect to varying training examples during the model exchange \cite{512}, e.g., continuous reconfiguration of their operating hash parameters \cite{513}, in the data block formulation, which helps achieve stable and reliable blockchain storage and operations.  }

By using blockchain, FLchain can attract more devices in the learning process for better scalability thanks to its decentralized network topology. Particularly, the information on learning updates 
is secure in a form of immutable blocks which improves the security of FL training in edge networks \cite{29}. With its decentralized and secure nature, {FLchain} is promising to provide attractive solutions for enabling intelligent and scalable edge networks with security guarantees. The comparison between \textcolor{black}{the classical FL and FLchain} is summarized in Table~\ref{Table:Comparison}. In addition to benefits, FLchain still remains some drawbacks such as possible latency from blockchain mining, resource management, and security from curious miners, which need to be considered in system designs. In this article, we perform a holistic investigation and discussion of related issues in FLchain design via some key use cases, which will be presented in detail in Section~\ref{Sec:BlockchainFL_Design}.

%% file: BlockchainFL_Design.tex
\section{FLchain in Edge Computing: Design and Use Cases}
\label{Sec:BlockchainFL_Design}
In this section, we present the design and key use cases of FLchain in edge computing based on the recent works related to FL-blockchain integration. We cover four key domains in FLchain design, including communication cost, resource allocation, incentive learning, \textcolor{black}{security and privacy protection}. In each domain, the problem statement is presented and some key use cases are discussed. \textcolor{black}{Then, the key lessons learned with the most important design features are summarized,} and the outlooks for further research are also highlighted. 

\subsection{Communication Cost}
\subsubsection{\textcolor{black}{Problem Statement}}
In FLchain, each client computes and exchanges its training updates via a blockchain ledger running on the top of edge networks to perform the global model aggregation at local devices without requiring a central server. \textcolor{black}{Although the network costs, e.g., latency, caused by communications with the central server are eliminated, the use of blockchain introduces new costs associated with block mining.} Therefore, the latency formulation for FLchain needs to take both on-device training latency, update communication latency, \textcolor{black}{and block mining latency into account with training accuracy awareness.}

\subsubsection{Use Cases}
\begin{figure}[t]
	\centering
	\includegraphics[width=1.00\linewidth]{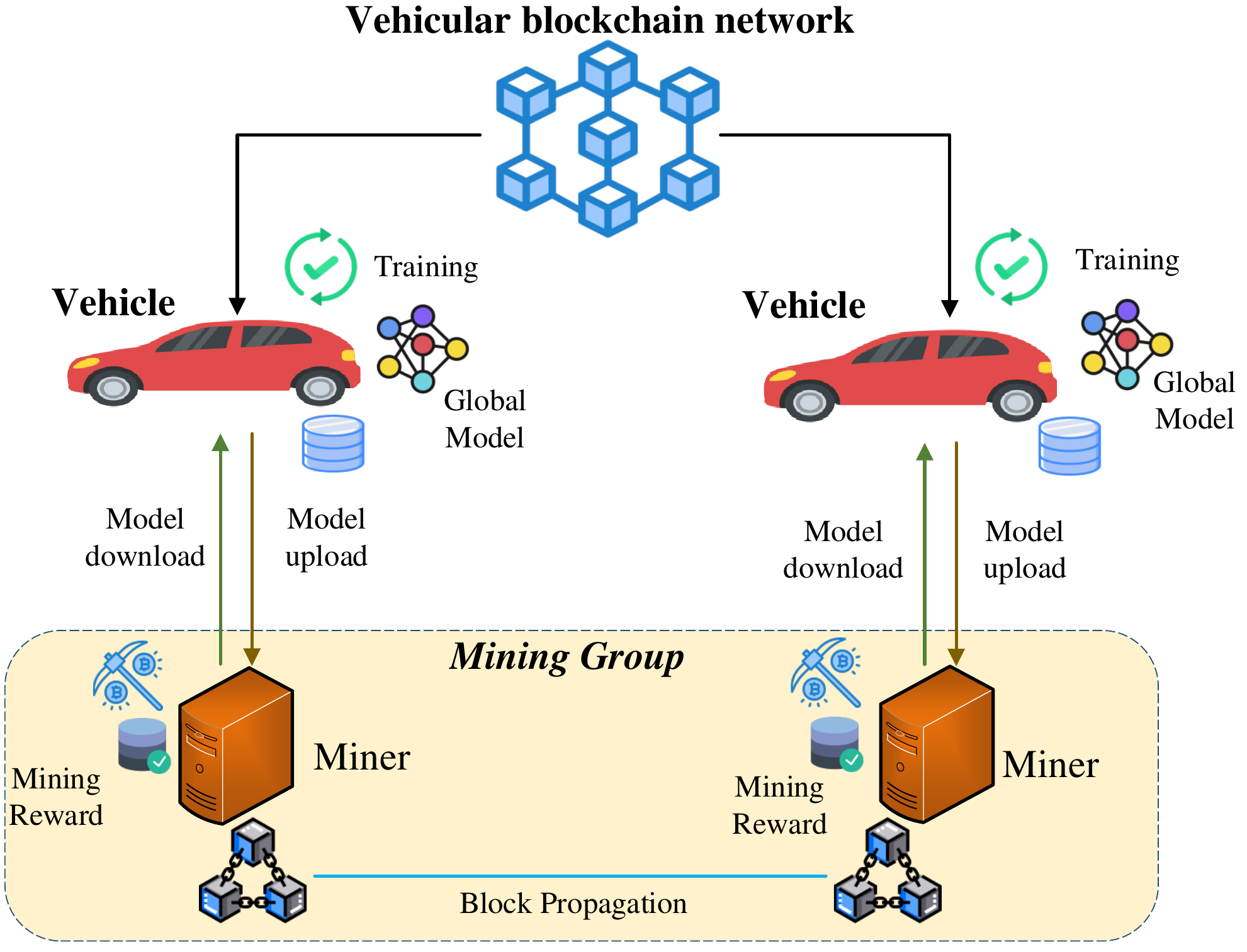}
	\caption{FLchain design for vehicular networks \cite{27}.}
	\label{Fig:FLchainDesign_Vehicular}
	\vspace{-0.1in}
\end{figure}

The work in \cite{25} is the first attempt to perform a holistic analysis of communication costs in FLchain. Here, MEC severs act as miners to mine blocks containing local updates, by performing a cross-verification on received local updates. To generate blocks, miners cooperate to run a PoW puzzle until they find a nonce value or receive a generated block from other miners. An integrated framework of computation, communication and block generation delays in each communication round is formulated by using an additive white Gaussian noise channel model in the communication phase and a block propagation delay model in the mining phase. The ultimate objective is to find an optimal block generation rate to minimize the average latency incurred by the PoW process while avoiding the possibility of forking caused by incorrect global updating in the block generation phase. Compared to classic FL approaches, such as Vanilla FL \cite{26}, FLchain achieves a lower system latency with a better learning accuracy in different simulation settings. Another study in \cite{27} performs a communication cost analysis for an 
\textcolor{black}{FLchain} 
in vehicular networks, as indicated in Fig.~\ref{Fig:FLchainDesign_Vehicular}. The use of blockchain potentially helps overcome the centralized malfunctioning problem and enhance the scalability of the intelligent vehicular network by attracting untrustworthy vehicles with a reward mechanism to improve the FL training performance. In this case, each vehicle acts as an FL client to perform local training and exchange with other vehicles via its miner in a blockchain network. More specifically, the latency for FL training and model updating at a vehicle $n$ with its associated miner is formulated by jointly considering model computation, model offloading, and blockchain mining as follows:
\begin{equation}
T_n = \frac{\tau_{out}+\mathcal{T}_g}{1-p_{fork}} +\tau_{global}+\tau_{gdn}, 
\end{equation}
Here, $\tau_{out}$ is timeout of local update dumping in the PoW process, $\mathcal{T}_g$ is the block arrival/generation delay, $p_{fork}$ is the forking probability, and $\tau_{global}$ and $\tau_{gdn}$ are the delays for global model update and global model downloading, respectively. The latency is then optimized by an online delay minimization algorithm with respect to transmission frame size, block size, and block arrival rate under different channel conditions. Simulation results reveal the benefits of the joint consideration of communication latency and consensus delay in achieving an optimal FLchain training latency, and the positive impacts of blockchain parameters, e.g., block arrival rate, on the overall model learning performance. However, the comparison of delay performances of the proposed FLchain and advanced FL schemes in edge computing \cite{28} has not been implemented. 

To minimize system delays for FLchain, a new edge association solution is proposed in \cite{29} for digital twin-based edge networks. Here, digital twin is integrated into wireless networks in a fashion that IoT devices are mapped with twins in MEC servers. This potentially reduces communication latency and enhances reliability for the edge computing plane due to the reduction of the impact caused by unreliable communications between users and MEC servers. A blockchain platform running on a Delegated Proof of Stake (DPoS) \cite{30} consensus mechanism is adopted to establish the decentralized training network for FLchain in edge computing. Accordingly, the latency of the digital twin-based FLchain is formulated as the sum cost of local training on digital twins, model aggregation on BSs, transmission of model parameters, and block verification. A multi-agent deep reinforcement learning (DRL) algorithm is then proposed to solve the latency optimization problem with respect to learning accuracy conditions and bandwidth allocation constraints. Simulations based on a CIFAR10 dataset consisting of 60,000 images in 10 classes are performed in an edge network with 5 BSs and 100 end users, showing that the proposed scheme can achieve a lower system latency without compromising training performances. 

\subsubsection{Lessons Learned and Outlook}
\begin{itemize}
	\item	Communication cost is a key issue to be solved before deploying FLchain systems in practice. 
    \textcolor{black}{Most of current FLchain schemes analyze communication costs by considering on-device training latency, latency of communication with miners, and mining latency \cite{25,27}, while the work in \cite{29} focuses on edge computing latency and parameter transfer communication latency.}
	\item	Some possible solutions can be adopted to achieve lower communication costs, such as reward-based training \cite{27} and adjustment of block arrival rate \cite{29} with respect to learning accuracy and bandwidth resource constraints. 
	\item	Most of current solutions \cite{25,27} rely on the PoW consensus mechanism, which often requires large bandwidth and energy resources to implement the mining process. Therefore, developing lightweight blockchains should be considered to facilitate FLchain \cite{31}, aiming to enhance the users' experience of training and reduce system costs.
\end{itemize}

\subsection{Resource Allocation}
\subsubsection{Problem Statement}
\textcolor{black}{Resource allocation is important to support the operations of FLchain. At the client side}, when devices need to share their computation and storage resources for both data training and blockchain participation. \textcolor{black}{At the MEC server side,} resource management is also needed to ensure efficient mining in a way blocks consisting of local updates are appended to the blockchain with \textcolor{black}{minimal forking probabilities}. 

\subsubsection{Use Cases}
The work in \cite{32} presents a DRL-based resource allocation strategy for FLchain in order to solve two critical issues in federated data training, including increased latency due to block mining and resource constraints of mobile devices for local training. To overcome these challenges, the update communication among servers (e.g., BS) and mobile devices is formulated as an $M/M/c$ queue model, where $M$ denotes the Poisson arrival and exponential service processes and $c$ is the number of serving nodes. For example, in the FLchain communication, new transactions are stored in a queue before being appended into a block, and the mining winner is the serving node which is responsible for running PoW puzzle. By using a Markov process, a reward function is derived in a fashion the energy consumption and training latency are minimized while the model accuracy rates are maximized, with respect to uplink and downlink bandwidth resource constraints. Experiments demonstrate the efficiency of the proposed DRL algorithm, which can reduce the energy consumption up to 72\% in comparison with a greedy scheme. DRL is also used in \cite{33} for resource allocation, by allowing model owners to find the optimal decisions on the energy and the channels to perform training without prior knowledge, with respect to the uncertainty of energy levels and user mobility. 
A stochastic optimization problem of the model owner is formulated to maximize the number of 
\textcolor{black}{communication rounds}
while minimizing the resource usage of energy and channel bandwidth. The use of an action policy allows model owners to determine which channel should be used to transmit global FL models to clients and estimate how much energy resource should be released to minimize system costs. Recently, a resource management scheme is considered to analyze the tradeoff between learning accuracy and resource consumption for FLchain with digital twin in MEC networks \cite{34}. A lightweight blockchain platform based on DPoS consensus is designed to support model updating and block mining in FLchain. A DRL approach is also adopted to implement the optimal user and resource scheduling to improve both communication efficiency and training performance, by solving a data relaying optimization problem with deep neural networks as a strategy scheduler for all IoT users. To evaluate the performance of the proposed FLchain scheme, real-world MNIST dataset and Fashion-MNIST dataset \cite{35} are used in a simulated environment with 100 end users and 4 BSs. Through simulations, DRL demonstrates its effectiveness in finding the optimal solutions for user scheduling and bandwidth allocation, while ensuring acceptable learning rates in various network settings. 
\subsubsection{Lessons Learned and Outlook}
\begin{itemize}
	\item 	Resource allocation is an important task to ensure optimal resource usage for FLchain data training. \textcolor{black}{Most of the state-of-the art techniques \cite{32}, \cite{33} use DRL to implement resource allocation strategies for FLchain systems under energy and channel bandwidth availability conditions.}
	\item	Lightweight blockchain platforms have been adopted based on DPoS consensus to support model updating and block mining in FLchain \cite{34}, which has the potential to mitigate energy consumption for FL training. 
	\item	Some possible directions can be considered to further improve resource allocation, by jointly considering block generation \cite{36}, model arrival rate, and on-device learning rate. In this way, we can achieve cooperative resource allocation solutions for both miners and local devices in FL-based intelligent edge networks. 
\end{itemize}

\subsection{Incentive Learning }
\subsubsection{Problem Statement}
Although FL has promoted collaborative learning while preserving user privacy, it still faces a challenge of incentivizing clients to participate in the FL process \textcolor{black}{and contribute their data as well as computation resources}. \textcolor{black}{Without a proper incentive  mechanism}, clients may not be willing to join the data training which would reduce the scalability of the designed FL system \cite{37}. Recently, blockchain has emerged as an attractive tool to \textcolor{black}{facilitate transparent economic mechanism designs, so as to boost} the FL training in FLchain systems. 
\subsubsection{Use Cases}
Some use cases towards incentive learning for FLchain have been presented in the literature. For example, the work in \cite{38} implements a rigorous reward policy design for FLchain, by providing an economic solution to realize desired objectives under the rational assumption of mobile users. The key objective is to offer repeated competition for FLchain so that rational users follow the protocol and maximize their economic profits. Each user chosen in a certain training round can select top model updates of other users in the previous round to update its model via blockchain. The users with the highest votes receive rewards for the next round of learning. The efficiency of the proposed incentive mechanism is investigated via novel auction theories such as contest theory with multiple prizes, showing a good incentive compatibility. An incentive mechanism for FLchain learning is also introduced in \cite{39}. Particularly, an integrated blockchain-based decentralized reputation system is designed to ensure trustworthy collaborative model training in edge computing environments. In this context, blockchain is extremely helpful to solve a wide range of issues such as data security, centrality of model training and centralized points-of-failures and compromises. To improve the performance of global model training, incentivization is highly necessary to recruit more mobile users with high-quality data sources. The work in \cite{40} considers an FLchain scheme for collaborative learning among data owners, cloud vendors, and AI developers in a trustless AI marketplace in blockchain-based cloud computing \cite{nguyen2019integration}, aiming at protecting the privacy and ownership of assets. Some system chaincode functions are configured to incentivize all clients to immutably record their actions in the distributed ledger so that blockchain can verify behaviours of users in FL training. A prototype is set up with three different organizations on a Hyperledger Fabric blockchain, each of them has 24 peers to perform shared AI model training via blockchain, showing a low network latency and improved throughput (e.g., number of transactions per second) under different network settings. 

A decentralized, public auditable FLchain system is designed in \cite{41} with trust and incentive to accelerate the cooperation in FL training without worrying about trust among entities. In fact, blockchain is promising to store securely information of trainers and FL models during the FL process, and mining helps to verify incorrect updates from FL computation. Particularly, any clients who detect any misbehaviour are rewarded and affected trainers are compensated through the blockchain-based incentive mechanism. Similarly, a value-driven incentive mechanism with blockchain is designed for FL, called DeepChain, aiming to encourage parties to behave correctly and share actively the obtained local gradients during the collaborative learning \cite{42}. Each party can get rewards or penalties based on his contribution. The benefits of learning incentivization are twofold, including attracting more clients to join the training and ensuring that clients are honest in local model training and update exchange.  Also, model updates are verified and audited by blockchain, while gradient collecting and parameter update are monitored by all participants over the ledger network, which thus ensures fairness in the FL process.  In addition to incentivization, FL servers need to pay clients as a contribution cost after the FL training. To solve this issue, a payment method is considered in \cite{43} by designing a Class-Sampled Validation-Error scheme to validate the gradients of clients to specify reasonable device rewards. A smart contract is integrated in a Hyperledger Fabric blockchain to implement gradient verification and reward scheduling in a transparent and trustful way \cite{517}. In this way, clients can benefit by retaining ownership and receiving incentives for their data, while model owners are able to access a larger dataset, which thus enhance the robustness of the overall FL learning. 
\subsubsection{Lessons Learned and Outlook}
\begin{itemize}
	\item Several literature works have been presented to design incentive mechanisms for FLchain systems, \textcolor{black}{with the common objective of attracting more clients and datasets in AI training and improving the robustness of the overall FLchain learning \cite{38}, \cite{39}, \cite{41}.} 
	\item	Blockchain has provided \textcolor{black}{unique features (e.g., decentralization,  trustworthiness) for enabling secure incentive solutions} in FL training where model updates are verified and audited by blockchain, while gradient collecting and parameter update are monitored by all participants over the ledger network, which thus ensures fairness in the FL process \cite{42}, \cite{43}.
	\item	However, the costs of verification for gradient updates during the incentive process have not been considered. In the future, it is important to analyze the tradeoff between reward profits and system costs, which benefits both clients and model owners, where auction theories can be useful for incentive analysis \cite{44}. 
\end{itemize}

\subsection{\textcolor{black}{Security and Privacy Protection}}
\subsubsection{Problem Statement}
To ensure the robustness and safety of FLchain systems, building \textcolor{black}{security and privacy protection mechanisms} is of paramount importance. Attacks such as poisoning attacks \cite{49} and data privacy threats can make FLchain systems vulnerable and risky especially in distributed edge networks consisting of multiple devices and servers. Thus, developing solutions to address security and privacy issues is highly needed before FLchain can be deployed at a large scale. 

\subsubsection{Use Cases}
Some use cases related to security protection for FLchain have been considered in recent works. The work in \cite{45} investigates an FL-block system where users participate in the collaborative learning and exchange their local updates via edge servers in a blockchain network, while data is stored in a distributed hash table. An attack model is taken into account; that is, an adversary can make attempt to train a local model using designed falsified data and replace the global model before update transmission, aiming to manipulate the training output. By adjusting the difficulty level of blockchain mining, the possibility of poisoning attacks on training data can be reduced, without degrading training performance. A blockchain-based decentralized secure multiparty learning system is studied in \cite{48}. Each client computes and broadcasts its local model and executes the received models from other clients via blockchain using its own dataset after calibration. Two types of Byzantine attacks in FLchain are considered at model broadcasting and model calibration processes, which can be solved by using a cooperative mining strategy with off-chain sample mining and on-chain mining. 

In terms of privacy protection for FLchain, some solutions have been also proposed. A blockchain-based FL architecture called PriModChain is introduced in \cite{46} for industrial IoT networks with a focus on privacy protection in model training and update transmission. A differential privacy approach is applied to locally generated models with artificial noise to reduce the possibility of individual record identification. The communication to exchange the global ML model between the central authority and distributed users can be secured by using smart contracts, which helps achieve an agreement on the update verification and provides transparency for the FL updating. This feature enforces unbiased and error-fee data manipulations, which thus enhances safety and reliability of the FL process against external data threats. 

{\color{black}It is noting that in the differential privacy-enabled FLchain, artificial noises aims to disturb the trained gradient by using a privacy mechanism such as Laplace, Duchi, or Piecewise one, before offloading to its associated MEC server. Then, all MEC servers incorporate all client's differentially privacy updates and perform consensus to generate an aggregated global gradient which is then broadcast to all participating clients for next round of training. After the global aggregation, a single node cannot reverse the ground truth to retrieve the actual vector containing client's updates  during the blockchained model exchange \cite{515}. Even in the public model update sharing between a client and its associated MEC server where this server performs multiple queries to try to retrieve the actual update of the client, the client is able to realize this unauthorized behaviour by tracing the blockchain ledger log or using the traceability of smart contracts deployed in the MEC network \cite{516}. }

In line with the discussion, the work in \cite{47} proposes a scalable privacy architecture for blockchain-based distributed learning by providing validation on the sources quality and confidentiality on the model within a privacy-preserved coalition among untrusted learning parties. Attacks can steal private data from a party during the training or try to reconstruct the training set from the generated gradients. In this context, differential privacy techniques can be employed so that model computation is considered to be differentially private if the output is independent to particular data points from input data. 
\begin{table*}
	\centering
	\caption{Taxonomy of FLchain design and use cases. }
	{\color{black}
		\label{Table:FLchaindesign}
		\begin{tabular}{|p{0.5cm}|p{0.4cm}|P{2.5cm}|P{1.2cm}|P{1.4cm}|P{4.4cm}|P{4.5cm}|}
			\hline
			\textbf{Issue}& 	
			\textbf{Ref.} &	
			\textbf{Use Case}&	
			\textbf{FL Clients}& 
			\textbf{Consensus/ Blockchain}& 	
			\textbf{Key Contributions}&
			\textbf{Limitations}
			\\
			\hline
			\parbox[t]{1.5cm}{\multirow{9}{*}{\rotatebox[origin=c]{90}{Communication cost}}} & 
			\cite{25} &	On-device FL &	Mobile devices &	PoW &	An end-to-end latency analysis for on-device FL with blockchain. &	Other parameters such as block arrival rate, rewards have not been considered in system analysis. 
			\\ \cline{2-7}&
			\cite{27}&	Vehicular blockchain-FL&	Vehicles&	PoW&	A comprehensive delay analysis for vehicular FLchain. &	The delay comparison between FLchain and classic FL schemes has not been conducted. 
			\\ \cline{2-7}&
			\cite{29}&	Digital twin-based FLchain & IoT devices&	DPoS&	\textcolor{black}{A communication analysis model for the edge computing plane in digital twin-based MEC networks with blockchain.} &	Comparison between DPoS with other consensus schemes has not been performed.
			\\ \cline{2-7}
			\hline
				\parbox[t]{1.5cm}{\multirow{8}{*}{\rotatebox[origin=c]{90}{Resource allocation}}} & 
			\cite{32}&		Resource management in FLchain training&	Mobile devices&	PoW&	A DRL-based approach for resource management in FLchain. &	Resource allocation for block mining at miners has not been considered.
			\\ \cline{2-7}&
			\cite{33}&	Resource management in FLchain training &	Mobile devices&	-&	A DRL-based approach for resource allocation in global model transmissions. &	The role of blockchain in FL training has not been verified. 
			\\ \cline{2-7}&
			\cite{34}&	Resource allocation for FLchain&	IoT devices&	DPoS&	A DRL-based strategy for optimal resource allocation in FLchain training. &	The impacts of blockchain on training has not been investigated. 		
			\\ \cline{2-7}
			\hline
			
			\parbox[t]{1.5cm}{\multirow{12}{*}{\rotatebox[origin=c]{90}{Incentive learning}}} & 
			\cite{38}&		Incentive FLchain design&	Mobile users&	PoW /Ethereum&	A reward policy design for incentive in FLchain&	Simulations for FLchain training have not been performed. 
			\\ \cline{2-7}&
			\cite{39}&	Incentive FLchain design&	Mobile users&	-&	A reputation scheme for FLchain training. &	Simulations for FLchain training have not been performed. 
			\\ \cline{2-7}&
			\cite{40}&	Incentive FLchain design&	Data owners&	Hyperledger Fabric&	An incentive protocol for supporting FLchain clients. &	Training performance of FL has not been evaluated.  
			\\ \cline{2-7}&
			\cite{41}&	Incentive mechanism for auditable FLchain&	Mobile users&
			-&	A decentralized, public auditable FLchain system with incentive. &	Incentive performance on blockchain has not been investigated.
			\\ \cline{2-7}&
			\cite{42}&	Incentive mechanism for FLchain&	Mobile users&	Ethereum&	A value-driven incentive mechanism with blockchain for FL. &	Security ability of proposed consensus scheme has not been verified. 
			\\ \cline{2-7}&
			\cite{43}&	Payment method for FLchain&	Mobile users&	Hyperledger Fabric&	A payment method for clients in FLchain training. &	Smart contract costs have not been simulated. 
			\\ \cline{2-7}
			\hline
			
			\parbox[t]{1.5cm}{\multirow{12}{*}{\rotatebox[origin=c]{90}{Security and privacy protection}}} & 
			\cite{45}&	Attack detection in FLchain&	Mobile users&	PoW&	A Poisoning attack detection and defense solution in FLchain with learning performance awareness. &	Numerical simulations for attack in blockchain mining are lacked. 	
			\\ \cline{2-7}&
			\cite{48} &	A secure model for FLchain training&	Mobile users&	PoW&	Blockchain-based decentralized multiparty learning system with high security protection. &	Complexity of the proposed scheme has not been evaluated. 
			\\ \cline{2-7}&
			\cite{46}&	Privacy protection in FLchain training &	IoT users&	Ethereum&	A privacy protection solution for model training and update transmission in FLchain. &	Attack simulations have not been provided. 
			\\ \cline{2-7}&
			\cite{47}&	Privacy preservation in distributed learning&	Data users &	Ethereum&	A scalable privacy architecture for blockchain-based distributed learning. &	System costs, such as latency, incurred from privacy preservation has not been considered. 
			\\ \cline{2-7}
			\hline
	\end{tabular}}
\end{table*}

\subsubsection{Lessons Learned and Outlook}
\begin{itemize}
	\item In FLchain, attacks can make attempt to train a local model using designed falsified data to replace the global model and modify parameter values during model transmission, aiming to manipulate the training output. \textcolor{black}{Most of existing works focus on building attack detection mechanisms for model training and update transmission in FLchain \cite{45,46,47}.}
	\item The possibility of surface attacks (e.g., poisoning ones on training data) can be reduced by adjusting mining difficulty without degrading training performance \cite{45}. Smart contracts are useful to build secure communication between the central authority and the distributed users for safe model updating \cite{46}.
	\item	\textcolor{black}{For privacy protection, differential privacy techniques can be employed in FLchain so that model computation is considered to be differentially private if the output is independent to particular data points from input data \cite{46}, \cite{47}. In this way, differential privacy is applied to locally generated models with artificial noise to reduce the possibility of individual record identification.}
	{\color{black}In the future, deploying more secure spatial decompositions algorithms is desired. For example, a solution is proposed in \cite{514} for indefeasible Laplace noise provision to address the problem of symmetric distribution of Laplace mechanisms in differential privacy. In this way, indefeasible noises are added to the randomly selected leaf child of each intermediate node of the private tree through multiplying Laplace noises, which helps prevent the possibility of noise cancellation and satisfy $\epsilon$-differential privacy. }
	
	\item	Moreover, attack models should be considered for both blockchain mining and local training in FLchain. For example, greedy miners may exploit the FLchain system by augmenting their mining power to make control on the miner group to modify data blocks, or double-spending issues with a Sybil attack can make mining inefficient \cite{50}. \textcolor{black}{Moreover, to enhance privacy in FLchain, some techniques such as homomorphic encryption \cite{add2} would be very useful for privacy-preserving outsourced storage and computation in FL training where data can be encrypted before sharing on the blockchain for FL model aggregation. This is extremely important in FLchain applications such as healthcare \cite{add3} in which health data and personal information are highly sensitive and privacy protection is highly needed.} 
\end{itemize}
In summary, we list FLchain design issues and use cases in Table~\ref{Table:FLchaindesign} to summarize the technical aspects as well as the key contributions and limitations of each reference.

\begin{figure*}[t]
	\centering
	\includegraphics[width=1.00\linewidth]{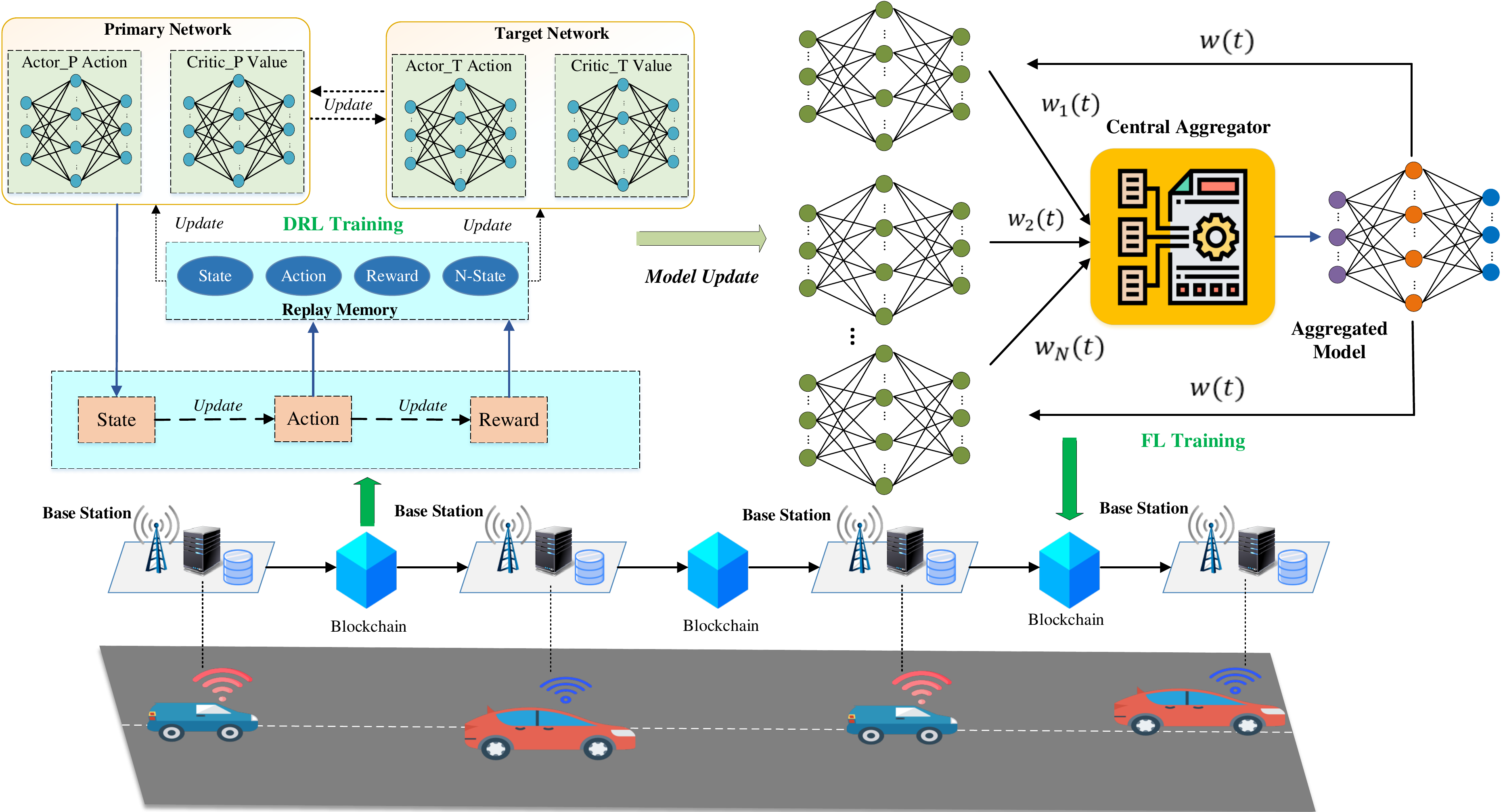}
	\caption{\textcolor{black}{The architecture of FLchain-based data sharing for IoV.} }
	\label{Fig:FL_DataSharing}
	\vspace{-0.1in}
\end{figure*}

%% file: BlockchainFL_Applications.tex
\section{FLchain Applications in Edge Computing}
\label{Sec:BlockchainFL_Applications}
In this section, we focus on analyzing the applications of Blockchain-FL 
in some popular applied domains, such as edge data sharing, edge content caching, and edge crowdsensing. 
\subsection{FLchain for Edge Data Sharing}
 \begin{figure*}
	\centering
	\includegraphics[width=0.6\linewidth]{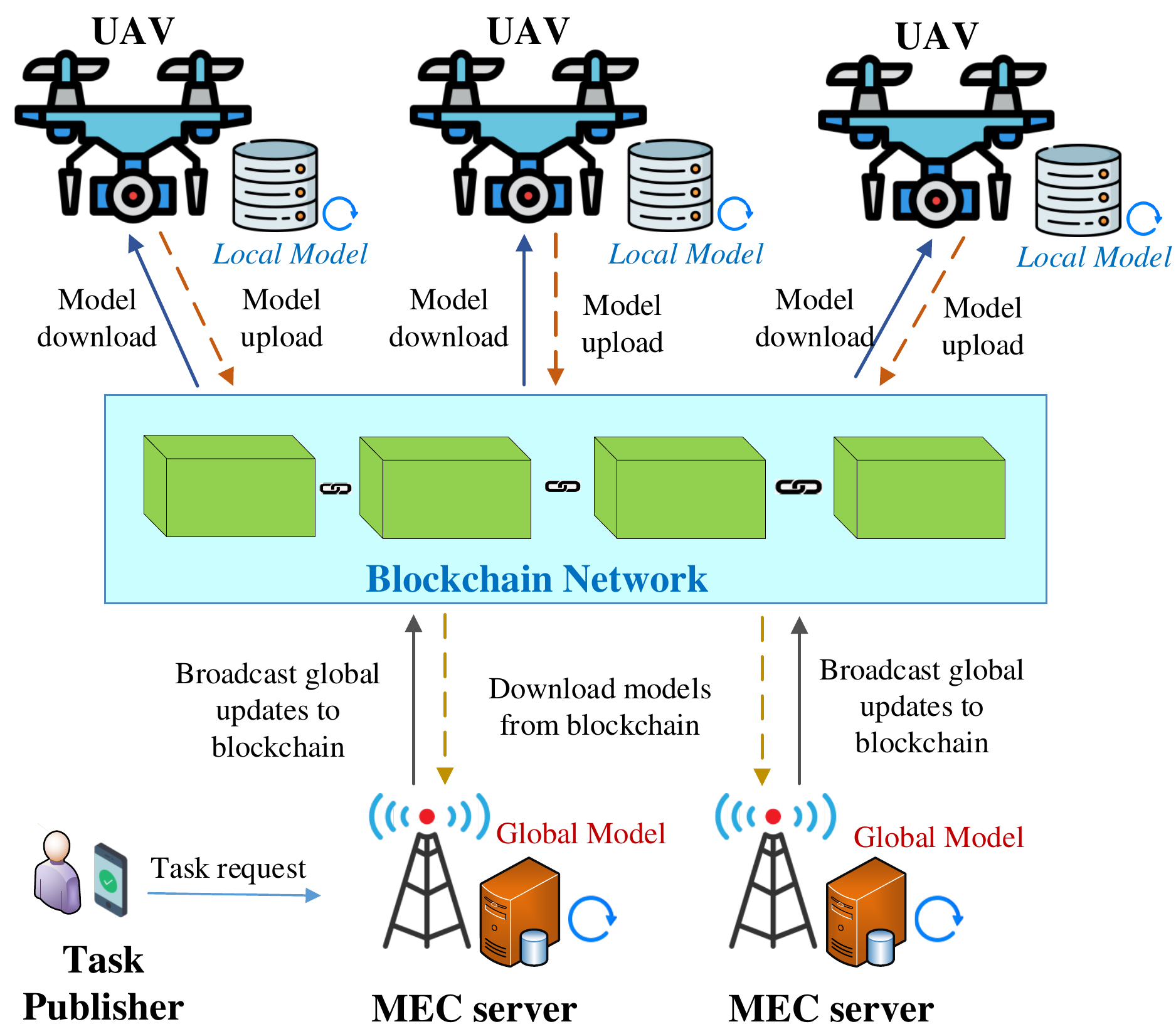}
	\caption{\textcolor{black}{FLchain for edge crowdsensing in UAV networks \cite{61}.}}
	\label{Fig:FL_Crowdsensing_UAV}
\end{figure*}
\textcolor{black}{IoT data sharing} is one of the key services in edge networks, aiming to transfer data over a shared environment to serve end users based on edge computing. In fifth-generation (5G) and beyond networks, providing reliable and scalable \textcolor{black}{IoT data sharing solutions} to meet the increasing data usage demands of users \cite{51} is highly important. FLchain with its distributed learning and security nature can facilitate edge data sharing. For instance, an FLchain scheme is proposed in \cite{52} for privacy-preserved \textcolor{black}{IoT data sharing} among distributed multiple parties such as mobile devices in industrial IoT networks with a BS. Given a data sharing request, a multiparty data retrieval process is executed to seek the related parties (or data owners) as committees based on registration records. These committee nodes act as miners to train the global model to output the training results and then return to the requester. They also perform mining of block consisting of data sharing transactions collected over a given period of time. As a result, data owners and data requestors can achieve reliable and fast data sharing while the sharing records are stored on the blockchain for tracing. \textcolor{black}{Another FLchain-based \textcolor{black}{vehicular data sharing} framework in edge computing is proposed for the Internet of Vehicles (IoV) in \cite{53}, as illustrated in Fig.~\ref{Fig:FL_DataSharing} with three key phases: node selection, local training, and global model computation. A node selection problem is formulated to select vehicles for local data training, which then generates model updates to build a global model. More specifically,} each vehicle acts as an FL client to cooperatively share data with an aggregation server at a macro BS (MBS). Vehicles with different service demands, e.g., traffic flow estimation or path selection can submit a data sharing request to the MBS. By running a shared global model using accumulated vehicular datasets from connected vehicles, the MBS transforms the data sharing process into a computation task, in order to handle the sharing requests from vehicles with a DRL algorithm for sharing cost minimization. Particularly, to provide the security and reliability of the vehicular data sharing, an immutable blockchain ledger is deployed on the top of the edge vehicular network to implement the verification of model parameter updates and store them in blocks in a decentralized manner. 


FL is also incorporated with a hierarchical blockchain to build a knowledge sharing in edge-based vehicular systems \cite{54}. The information of trained parameters can be shared over the IoV network consisting of RSUs and BSs which can be classified into separate groups based on their regional features under the management of a local blockchain running on a light-weight Proof-of-Knowledge (PoK) consensus mechanism. To realize intelligent distributed learning, an FL process is carried out among vehicles and RSUs in a certain area before sharing the trained model with a global FL network on the RUSs-BSs network. To achieve a secure \textcolor{black}{IoT data collaboration} among IoT users, an FLchain framework is also studied in \cite{55} among private data center, a public data center, and a blockchain network. Here, the data center aims to collect data from IoT environments based on FL which can exploit the computational resources of IoT devices to build a powerful model. Models with parameters are then stored in a distributed data storage system at the public data center with cryptography protection. To provide flexible and secure data management, blockchain is adopted based on a smart contract design, which can support access control over the data usage among users. Simulations with human movement datasets confirm high performance in training accuracy and high-security degrees for a wearable sensor network. 

\subsection{FLchain for Edge Content Caching}
Due to the high data distribution of large-scale mobile edge networks, FLchain can provide distributed AI solutions to support intelligent and secure content caching at the network edge \cite{56}. Indeed, data offloaded from IoT devices can be cached by MEC servers, where FL plays the role of creating intelligent caching policies to cope with the explosive growth of mobile data in modern IoT networks. FL can help overcome the challenges faced by traditional centralized learning approaches in terms of high privacy concerns since the fact that mobile users may not fully trust third-party servers and thus hesitate to share their private data to MEC servers. Meanwhile, blockchain would be useful to build trust between users and MEC servers for reliable content caching without the need for extra network management infrastructure. As shown in \cite{57}, FL can be integrated with blockchain to provide secure learning schemes for edge content caching in MEC networks. In this case, IoT devices can collaboratively train a neural network using their own datasets and then migrate the computed models to MEC servers for global aggregation. The ultimate goal is to estimate popular files to improve the overall cache hit rate at the network edge. To solve the issue of high communication latency caused by FL training, a gradient compression method is adopted to reduce the size of offloaded parameters. Then, a blockchain-based smart contract is integrated to establish a trustful training verification protocol among MEC servers and IoT users during FL communication rounds. Simulations from MovieLens datasets \cite{58} show a significant improvement in terms of high cache hit rate over the traditional FL-based caching algorithms. FL is also useful for enabling intelligent edge data caching by supporting distributed DRL training for content replacement in MEC networks. Mobile users can learn and train a shared model locally without sharing raw data to the network, while a cloud at the BS collects updates from distributed users to run a common model by averaging local updates. Based on the cooperation of users and cloud, a content replacement is formulated as a Markov decision process that is then solved by a DRL algorithm to optimize the cache hit rate with learning accuracy preservation. FL is adopted in \cite{59} to train DRL agents to provide guidelines for intelligent content caching at MEC servers in a fashion communication and computation costs in MEC systems are optimized. In the considered scenario, end users rely on an FL process to implement data DRL training directly without offloading raw data with personal information to MEC servers for privacy guarantees. Multiple MEC servers then coordinate to build a global DRL model, aiming to minimize system cost and achieve spectrum resource savings across the edge network. 

\subsection{FLchain for Edge Crowdsensing}
Mobile crowdsensing systems are designed to take advantage of ubiquitous mobile devices for sensing and collecting data from physical environments to implement data analytic tasks. Traditionally, centralized AI/ML architectures are often used to realize intelligent mobile crowdsensing \cite{60}, but they require direct access to sensing data which raises privacy concerns. Moreover, the utilization of a single server to handle all sensed data is not efficient especially when data are often distributed across a large-scale network. FLchain not only enables highly scalable intelligent edge data crowdsensing solutions but also provides high degrees of privacy and security based on FL and blockchain integration. For example, an FLchain scheme is studied in \cite{61} to construct a mobile crowdsensing system based on Unmanned Aerial Vehicles (UAVs), as indicated in Fig.~\ref{Fig:FL_Crowdsensing_UAV}. Blockchain is employed to build a decentralized learning network by interconnecting UAVs with data task publishers through an immutable ledger, which ensures that data training and model exchange among UAVs are monitored and traced for attack detection and data modification prevention.

Although FL can preserve privacy for users in AI training, data features embedded in model updates can be leaked which thus can reveal private user information \cite{62}. To overcome this challenge, a differential privacy technique is adopted by injecting artificial noise into the local gradient training in each communication round. Then, an incentive mechanism is integrated to attract more UAVs in the AI training to improve the overall FL performance. This approach is validated by a simulation based on the MNIST dataset including 60,000 training samples and 10,000 test samples with convolutional neural networks 
as the learning model. Evaluation results confirm a high utility of UAVs and low aggregation error while reducing convergence latency. Further, an incentive-based crowdsensing/crowdsourcing architecture is designed in \cite{63} based on FL for IoT in edge computing. Each client can receive an incentive after successfully uploading its computed parameter to the MEC server, in order to attract more users to contribute computational resources for running AI models. 
This incentive process is modeled by a two-stage Stackelberg game that shows advantages over the heuristic approach in terms of a utility gain enhancement by 22\% for different system settings. 
Another edge crowdsourcing scheme is investigated in \cite{64} based on an integrated FL-blockchain architecture in IoT environments. More specifically, a hierarchical crowdsourcing-empowered FL system is built for enabling distributed ML training at the network edge. The ultimate goal is to optimize the service quality of appliance manufacturers while preserving privacy with different privacy techniques. For security guarantees, blockchain is integrated with FL training to detect and prevent malicious attacks from modifying gradient updates and audit the parameter update of FL clients such as IoT devices. Implementation results prove the feasibility of the proposed FLchain with high training accuracy, security protection, and low system latency. 

In summary, we list FLchain applications in edge computing in Table~\ref{Table:FLchain_Applications} to summarize the technical aspects as well as the key contributions and limitations of each reference.

\begin{table*}[]
	\centering
	\caption{Taxonomy of FLchain applications in edge computing.  }
	{\color{black}
		\label{Table:FLchain_Applications}
		\begin{tabular}{|p{0.5cm}|p{0.4cm}|P{2.5cm}|P{1.2cm}|P{1.4cm}|P{4.4cm}|P{4.5cm}|}
			\hline
			\textbf{Issue}& 	
			\textbf{Ref.} &	
			\textbf{Use Case}&	
			\textbf{FL Clients}& 
			\textbf{Consensus/ Blockchain}& 	
			\textbf{Key Contributions}&
			\textbf{Limitations}
			\\
			\hline
			\parbox[t]{1.5cm}{\multirow{10}{*}{\rotatebox[origin=c]{90}{Data sharing}}} & 
			\cite{52} &	Industrial IoT data sharing&	IoT devices&	Proof of Training Quality&	A data sharing framework with privacy protection for industrial IoT with FLchain. &	Communication cost for proposed FLchain operations is high.
			\\ \cline{2-7}&
			\cite{53}&	Vehicular data sharing&	Vehicles&	DPoS&	A secure asynchronous FLchain-based data sharing model for IoV. &	Security performance has not been simulated. 
			\\ \cline{2-7}&
			\cite{54}&	Vehicular knowledge sharing&	Vehicles&	Proof-of-Knowledge &	A hierarchical FLchain scheme for vehicular knowledge sharing. &	Impacts of bloclchain consensus on FL sharing have not been investigated.
			\\ \cline{2-7}&
			\cite{55}&	Secure data collaboration& 	IoT users&	-	&	Secure data collaboration among IoT users with FLchain. &	Smart contract and mining performances have not been validated. 
			\\ \cline{2-7}
			\hline
			\parbox[t]{1.5cm}{\multirow{7}{*}{\rotatebox[origin=c]{90}{Content caching}}} & 
			\cite{57}&	Edge content caching&	IoT devices&	PoS&	An edge content caching framework for IoT networks with FLchain. &	Analysis of mining latency has not been considered.
			\\ \cline{2-7}&
			\cite{58}&	Edge content caching&	IoT users&	-&	An edge content caching scheme in edge networks with FL. &	Security solutions, e.g., blockchain, has not been integrated. 
			\\ \cline{2-7}&
			\cite{59}&	Edge data caching&	IoT users&	-&	A federated edge caching scheme based on DRL and FL. &	Performance of the scalability for the proposed FL model has not been shown.
			\\ \cline{2-7}
			\hline
			\parbox[t]{1.5cm}{\multirow{6}{*}{\rotatebox[origin=c]{90}{Crowdsensing}}} & 
			\cite{61}&	Edge crowdsensing&	UAVs&	PoW&	An FLchain scheme for mobile crowdsensing in UAV networks. &	Blockchain consensus and its impacts need to be evaluated. 
			\\ \cline{2-7}&
			\cite{63}&	Edge crowdsensing&	IoT users&	-&	A reputation-based crowd sensing/sourcing architecture. &	Blockchain performance has not been verified. 
			\\ \cline{2-7}&
			\cite{64}&	Edge crowdsourcing&	IoT devices&	PoS&	An FLchain for optimizing the service quality of appliance manufacturers. &	Training complexity has not been analyzed. 
			\\ \cline{2-7}
			\hline
	\end{tabular}}
\end{table*}

%% file: Challenges_Future-Directions.tex
\section{Unique Challenges and Future Directions}
\label{Sec:Challenges_Future-Directions}
As discussed in the above sections, {FLchain} demonstrates its increasingly significant role in supporting edge computing services and applications. Despite its great potential, {in this section, we would like to discuss unique research challenges} that need to be considered for future FLchain implementation in edge networks. {Here, we focus on a few unique challenges in FLchain, including security issues, communication and learning convergence issues, economic issues, plagiarism issues, and latency requirements of edge computing systems.} A number of possible research directions are also highlighted. 

\subsection{Security Issues in FLchain} 
Although blockchain can provide decentralization and traceability for FL training in edge computing, recent studies have found that blockchain still has its own security issues. For example, 51\% attack is a critical security problem in blockchain platforms \cite{65} where a group of powerful miners can take control of the mining with more than 50\% of the mining power of the network. These miners can modify block data or even prevent blocks from appending to the blockchain, which interrupts the communication among users (e.g., FL clients) and FL owners (e.g., MEC servers). {Another security issue is forking attacks \cite{challenge7} wherein \textcolor{black}{divergent blockchains are generated which results in a change in the block update rule and thus invalidates the synchronized block verification in blockchain. Such a forking attack} can mislead the blockchain nodes in the FL training process. Specifically, certain nodes can use different global parameters to train the shared FL model in some iterations, leading to incorrect FL model aggregation.} Other attacks such as double spending attacks and reentrancy attacks on smart contracts \cite{66} are issues to be solved before integrating blockchain in FL systems. {The fake parameter update from an adversary is also a security bottleneck \cite{zhanggan2020} \textcolor{black}{which can impersonate a certain client and broadcast fake parameters}. Such an adversary may deliberately broadcast fake samples where the miners cannot recognize during the mining process which leads to the incorrectness in the model aggregation.} 

To overcome these challenges, a solution is proposed in \cite{67} using a mining pool strategy to enhance the efficiency of the mining process, by solving security bottlenecks such as 51\% vulnerability with reduced block and transaction propagation delays. {However, the mining pool centralization may result in challenging selfish behaviors which poses threats to the system throughput.} Another promising solution is developing defense mechanisms for fighting against data threats in the consensus process within the blockchain network. Moreover, to mitigate the forking probability in blockchain, a large deviation theory is employed in \cite{challenge7} to analyze the vulnerability of blockchain networks incurred by intentional forks and propose strategic planning mechanisms for preventing the forking attacks. {In the future, it is necessary to take resource usage factors such as computational power and attack detection latency into the forking analysis.} {Furthermore, smart contracts can be also integrated in the miner network to perform reliable authentication for model updates from distributed nodes. By using self-executing contracts, all model update information is verified to evaluate the correctness of the message and user identification is authenticated for detecting possible fake updates.}

{\color{black}\subsection{Communication and Heterogeneity Issues in FLchain} 
Another challenge is the impact of the delay and errors of both communications and blockchain on the FL training performance. In fact, communications in FL training in both uplinks and downlinks are highly sensitive due to the unbalanced and non-IID data since the local training data at each client is different in size and distribution owing to different sensing environments. Further, when the number of clients grows exponentially, direct communications between numerous clients and a server for parameter updating become a bottleneck due to the increased traffic congestion in the network channels. This potentially incurs high delays and higher probabilities of data loss during the communication rounds, which results in slow and inaccurate FL training convergence rates. A possible solution to mitigate network traffic in the FL training is to use compression methods \cite{500} to reduce the size of the model update at each client at a communication round. However, such compression methods usually add to the error floor of the training objective as they increase the variance of the updates \cite{501}. Therefore, it is necessary to choose an appropriate number of quantization levels to strike the best error-communication trade-off. From the blockchain aspect, the latency and update bias caused by the block mining during the consensus also potentially degrade the performances of FL training, e.g., biased global model aggregation, which calls for urgent solutions to an efficient and reliable data learning in the FLchain system. 

In addition to that, the convergence of FL algorithms in FLchain systems is not always ensured due to the heterogeneous training capabilities of different IoT devices. Indeed, the characteristics of different IoT devices are different, ranging from hardware (CPU, memory), network protocols (e.g., 4G/5G and WiFi) to battery, all of which contribute to the heterogeneity of computation and communication of IoT devices and directly affect the one-device AI model training time and learning quality \cite{502}. For example, devices with limited computing capabilities and poor connection can take much longer to complete their local training and model update to the server, which can lead to serious delay in global aggregation. A direct application of existing FL protocols without any consideration of such heterogeneous device properties thus can make the overall training process inefficient. 

Several possible solutions have been proposed to solve issues related to FLchain performances. The research in \cite{8} proposes a new efficient communication protocol that is able to compress uplink and downlink communications while remaining high robustness to the increased number of clients, and data distribution. These properties can be achieved by using a combination of sparsification, ternarization, error accumulation, and optimal Golomb encoding techniques for uplink compression and speeding up parallel training in the global server without compromising learning convergence. Moreover, a new optimization algorithm is proposed in \cite{503} for FL-based IoT networks, called FetchSGD, that can train high-quality models for communication efficiency. At each communication round, clients compute a gradient based on their local data, then compress the gradient using a data structure called a Count Sketch before sending it to the central aggregator. The aggregator maintains momentum and error accumulation count sketches, and the weight update applied at each round is extracted from the error accumulation sketch. In this way, the proposed scheme reduces the amount of communication required each round, while still conforming to the training quality requirements of the federated setting. Further, designing lightweight consensus mechanisms to mitigate block mining delays is desired to boost the overall FLchain training speed, which can be done by implementing lightweight block validation schemes \cite{31}. The most important data such as local client updates can be prioritized to add to the block for the consensus among clients in the FLchain update aggregation, while signature assignment and block verification can be simplified with respect to the reputation of each client. For example, the client with higher reputation due to frequent update training contribution can be given a higher priority in the verification stage for fast block approval, which helps mitigate the delay and bias in the model aggregation and in turn improves the overall FLchain training performances. Further, to solve the heterogeneous issues in the FL training, the work in \cite{504} proposes a robust FL framework in heterogeneous environments where device clients may behave abnormally due to its behaviours and data points at different devices have different distributions. A modular algorithm is designed with three steps, including computing empirical risk minimizers, computing outlier-robust clustering algorithm, and implementing a distributed optimization on each cluster, aiming to solve the high dimension issues from heterogeneous behaviours of different devices in each network group.

}
\subsection{{Economic Issues in FLchain}}
In practical FLchain systems, when a mobile user serves as both training nodes and miners, how to encourage users to join the FLchain process is a key challenge. \textcolor{black}{A user may not be willing} to devote its resources to perform mining if it does not have much economic benefits \textcolor{black}{to compensate the consumption of} computational and storage resources \cite{challenge1, challenge2}. This issue is particularly important in server-less scenarios where there is no any servers (e.g., MEC servers) to coordinate the FLchain process and manage the blockchain mining. {Moreover, the heterogeneity among FLchain nodes caused by the different characteristics (e.g., CPU and memory) of devices can also result in imbalance among nodes \cite{70}. Therefore, the clients with better computational capability can dominate the training and mining to obtain more rewards. How to achieve an economic fairness among nodes in FLchain thus is a challenging issue.}

\textcolor{black}{A few solutions} have been proposed to solve economic issues in FLchain. For example, developing well-balanced incentive mechanisms for improving economic benefits is highly needed to encourage more users to implement FL training and blockchain mining. The authors in \cite{challenge3} implement a new reputation-based solution for the PoW computation in the blockchain, where mobile miners are incentivized to perform honest mining. A reputation-based algorithm is designed using a game theory to encourage users to join the mining process for extra profits with resource management awareness. Another work in \cite{challenge4} proposes a credit based incentive approach based on a revenue payment function of reputation for reward and punishment during the mining process. A cooperative mining behaviour is rewarded, while a non-cooperative behaviour with potential mining degradation is punished from the consensus process. This solution is promising to solve incentive issues in current FLchain networks, aiming to attract more users to join the mining process which in return enhances the robustness of the FLchain system. {However, a well-balanced incentive mechanism affects the optimization of computation resources at the client side. For example, a client might allocate 64\% CPU for FL and 36\% CPU for mining to maximize its profit. \textcolor{black}{Such an optimal allocation varies} with the incentive mechanism, which leads to different outcomes of AI model quality and learning latency. }

\subsection{{{Plagiarism Issues in FLchain}}}
{{Another unique challenge in FLchain is the plagiarism issue  during the block verification process, where  a lazy node may plagiarize the other clients' ML models without real training \cite{challenge5}.} Consequently, the lazy node can allocate more computational resources for the mining to gain more rewards. This issue not only causes unfairness among learning clients but also significantly reduces the overall FL training performance. Without efficient solutions for this issue, more clients will refuse to participate in the training process, which degrades the FLchain system.}

\textcolor{black}{A number of interesting directions} for solving lazy issues in FLchain may include the use of encryption techniques and incentive mechanisms. For example, a lightweight authenticated-encryption solution is proposed in \cite{challenge6} to provide reliable data encryption while saving communication bandwidth and memory resources for hardware-constrained IoT devices. Another solution in \cite{challenge4} uses incentive mechanisms with punishment policies applied to lazy nodes that copy trained parameters from other clients without performing local learning, while giving rewards to honest nodes for their training effort. This solution potentially prevents plagiarism issues in FLchain and encourages more clients to join the FL process for better training performance.

\subsection{Stringent Latency Requirements of MEC Systems} 
Implementing FLchain in MEC systems must meet the stringent latency requirement of customer services, such as edge-based autonomous driving or real-time healthcare analytics. FL can reduce communication latency due to optimized training without the need for raw data offloading, but it still introduces another latency issue from repeated communication rounds, which limits the convergence rate of an FL process. Moreover, the use of blockchain also introduces additional latency from block mining which poses new challenges for FL systems since an FL client needs to wait for the mining process to be completed before receiving model updates and performing its next training round. 

Therefore, some directions need to be taken to minimize the latency in FLchain implementation in edge computing. 
For example, an FL architecture is proposed in \cite{68} to accelerate the convergence speed of the FL training, by performing resource allocation in a fashion more resources should be allocated to devices with worse channel conditions or weaker computation capabilities. In this way, devices can achieve a balance in AI training power, thus improving the model aggregation rate for low-latency FL training. Moreover, light-weight mining designs should be focused, by reducing complexity in transaction formulation and block verification stages while security and scalability are preserved \cite{69}.
{\color{black}Moreover, to optimize the wireless communication latency caused by the model data offloading in the FLchain training, AI techniques such as DRL would be very useful to build latency control policies. For example, a DRL-based approach provided in \cite{509} can help client devices learn the optimal offloading rule with respect to update sizes, transmit power, and blockchain transaction states, aiming to minimize the communication delays while user privacy is ensured. Another DRL-based approach is also proposed in \cite{510} to schedule the sub-channel assignment and transmission power control for improving the wireless data transmission rates. In this context, both the strict reliability and latency requirements of all devices are taken into account in a multi-agent setting where mobile devices collaborate to optimize the  transmission latency based on local observation information. }

%% file: Conclusion.tex
\section{Conclusions} 
\label{Sec:Conclusion}
In this article, we have presented an overview on FLchain, an emerging paradigm in MEC enabled by the integration of FL and blockchain. \textcolor{black}{We have first introduced a generic FLchain architecture,} which opens a new fundamental way for enabling scalable and secure edge intelligence in next-generation wireless networks, \textcolor{black}{followed by the key concepts of FL and blockchain}. To realize FLchain in edge computing, we have analyzed some important design issues along with possible solutions via four use cases, including  communication cost, resource allocation, incentive learning, security and privacy protection. The lessons learned from the survey and outlooks have been provided. Furthermore, we have discussed the use of FLchain for some popular applications in edge networks, including edge data sharing, edge content caching and edge crowdsensing. Finally, we have outlined some key research challenges and possible directions toward the full realization of FLchain. 

%% file: FLchain_Survey_Manuscript.bbl
\begin{thebibliography}{10}
\providecommand{\url}[1]{#1}
\csname url@samestyle\endcsname
\providecommand{\newblock}{\relax}
\providecommand{\bibinfo}[2]{#2}
\providecommand{\BIBentrySTDinterwordspacing}{\spaceskip=0pt\relax}
\providecommand{\BIBentryALTinterwordstretchfactor}{4}
\providecommand{\BIBentryALTinterwordspacing}{\spaceskip=\fontdimen2\font plus
\BIBentryALTinterwordstretchfactor\fontdimen3\font minus
  \fontdimen4\font\relax}
\providecommand{\BIBforeignlanguage}[2]{{%
\expandafter\ifx\csname l@#1\endcsname\relax
\typeout{** WARNING: IEEEtran.bst: No hyphenation pattern has been}%
\typeout{** loaded for the language `#1'. Using the pattern for}%
\typeout{** the default language instead.}%
\else
\language=\csname l@#1\endcsname
\fi
#2}}
\providecommand{\BIBdecl}{\relax}
\BIBdecl

\bibitem{1}
A.~Al-Fuqaha, M.~Guizani, M.~Mohammadi, M.~Aledhari, and M.~Ayyash, ``Internet
  of {Things}: {A} {Survey} on {Enabling} {Technologies}, {Protocols}, and
  {Applications},'' \emph{IEEE Communications Surveys Tutorials}, vol.~17,
  no.~4, pp. 2347--2376, Fourthquarter 2015.

\bibitem{9200330}
D.~C. {Nguyen}, P.~{Cheng}, M.~{Ding}, D.~{Lopez-Perez}, P.~N. {Pathirana},
  J.~{Li}, A.~{Seneviratne}, Y.~{Li}, and H.~V. {Poor}, ``{Enabling} {AI} in
  {Future} {Wireless} {Networks}: {A} {Data} {Life} {Cycle} {Perspective},''
  \emph{IEEE Communications Surveys \& Tutorials}, Sep. 2020.

\bibitem{2}
H.~Li, K.~Ota, and M.~Dong, ``Learning {IoT} in {Edge}: {Deep} {Learning} for
  the {Internet} of {Things} with {Edge} {Computing},'' \emph{IEEE Network},
  vol.~32, no.~1, pp. 96--101, Jan. 2018.

\bibitem{3}
S.~Wang, T.~Tuor, T.~Salonidis, K.~K. Leung, C.~Makaya, T.~He, and K.~Chan,
  ``Adaptive {Federated} {Learning} in {Resource} {Constrained} {Edge}
  {Computing} {Systems},'' \emph{IEEE Journal on Selected Areas in
  Communications}, vol.~37, no.~6, pp. 1205--1221, Jun. 2019.

\bibitem{4}
C.~Ma, J.~Li, M.~Ding, H.~H. Yang, F.~Shu, T.~Q.~S. Quek, and H.~V. Poor, ``On
  {Safeguarding} {Privacy} and {Security} in the {Framework} of {Federated}
  {Learning},'' \emph{IEEE Network}, vol.~34, no.~4, pp. 242--248, Jul. 2020.

\bibitem{507}
D.~C. Nguyen, P.~N. Pathirana, M.~Ding, and A.~Seneviratne, ``{BEdgeHealth}:
  {A} {Decentralized} {Architecture} for {Edge}-based {IoMT} {Networks} {Using}
  {Blockchain},'' \emph{IEEE Internet of Things Journal}, 2021.

\bibitem{6}
U.~Majeed and C.~S. Hong, ``{FLchain}: {Federated} {Learning} via {MEC}-enabled
  {Blockchain} {Network},'' in \emph{2019 20th {Asia}-{Pacific} {Network}
  {Operations} and {Management} {Symposium} ({APNOMS})}, Sep. 2019, pp. 1--4.

\bibitem{7}
Y.~J. Kim and C.~S. Hong, ``Blockchain-based {Node}-aware {Dynamic} {Weighting}
  {Methods} for {Improving} {Federated} {Learning} {Performance},'' in
  \emph{2019 20th {Asia}-{Pacific} {Network} {Operations} and {Management}
  {Symposium} ({APNOMS})}, Sep. 2019, pp. 1--4.

\bibitem{9}
Q.~Yang, Y.~Liu, T.~Chen, and Y.~Tong, ``Federated {Machine} {Learning}:
  {Concept} and {Applications},'' \emph{ACM Transactions on Intelligent Systems
  and Technology}, vol.~10, no.~2, pp. 1--19, Jan. 2019.

\bibitem{10}
J.~Park, S.~Samarakoon, A.~Elgabli, J.~Kim, M.~Bennis, S.-L. Kim, and
  M.~Debbah, ``Communication-{Efficient} and {Distributed} {Learning} {Over}
  {Wireless} {Networks}: {Principles} and {Applications},''
  \emph{arXiv:2008.02608}, Aug. 2020.

\bibitem{11}
S.~Niknam, H.~S. Dhillon, and J.~H. Reed, ``Federated {Learning} for {Wireless}
  {Communications}: {Motivation}, {Opportunities}, and {Challenges},''
  \emph{IEEE Communications Magazine}, vol.~58, no.~6, pp. 46--51, Jun. 2020.

\bibitem{12}
W.~Y.~B. Lim, N.~C. Luong, D.~T. Hoang, Y.~Jiao, Y.-C. Liang, Q.~Yang,
  D.~Niyato, and C.~Miao, ``Federated {Learning} in {Mobile} {Edge} {Networks}:
  {A} {Comprehensive} {Survey},'' \emph{IEEE Communications Surveys Tutorials},
  vol.~22, no.~3, pp. 2031--2063, Thirdquarter 2020.

\bibitem{13}
Z.~Zhao, C.~Feng, H.~H. Yang, and X.~Luo, ``Federated-{Learning}-{Enabled}
  {Intelligent} {Fog} {Radio} {Access} {Networks}: {Fundamental} {Theory},
  {Key} {Techniques}, and {Future} {Trends},'' \emph{IEEE Wireless
  Communications}, vol.~27, no.~2, pp. 22--28, Apr. 2020.

\bibitem{14}
C.~Briggs, Z.~Fan, and P.~Andras, ``A {Review} of {Privacy} {Preserving}
  {Federated} {Learning} for {Private} {IoT} {Analytics},''
  \emph{arXiv:2004.11794}, Apr. 2020.

\bibitem{15}
R.~Yang, F.~R. Yu, P.~Si, Z.~Yang, and Y.~Zhang, ``Integrated {Blockchain} and
  {Edge} {Computing} {Systems}: {A} {Survey}, {Some} {Research} {Issues} and
  {Challenges},'' \emph{IEEE Communications Surveys Tutorials}, vol.~21, no.~2,
  pp. 1508--1532, Secondquarter 2019.

\bibitem{16}
J.~P. Queralta and T.~Westerlund, ``Blockchain for {Mobile} {Edge} {Computing}:
  {Consensus} {Mechanisms} and {Scalability},'' \emph{arXiv:2006.07578}, Jun.
  2020.

\bibitem{17}
J.~Konečný, H.~B. McMahan, F.~X. Yu, P.~Richtárik, A.~T. Suresh, and
  D.~Bacon, ``Federated {Learning}: {Strategies} for {Improving}
  {Communication} {Efficiency},'' \emph{arXiv:1610.05492}, Oct. 2017.

\bibitem{18}
T.~Li, A.~K. Sahu, A.~Talwalkar, and V.~Smith, ``Federated {Learning}:
  {Challenges}, {Methods}, and {Future} {Directions},'' \emph{IEEE Signal
  Processing Magazine}, vol.~37, no.~3, pp. 50--60, May 2020.

\bibitem{19}
Z.~Du, C.~Wu, T.~Yoshinaga, K.-L.~A. Yau, Y.~Ji, and J.~Li, ``Federated
  {Learning} for {Vehicular} {Internet} of {Things}: {Recent} {Advances} and
  {Open} {Issues},'' \emph{IEEE Open Journal of the Computer Society}, vol.~1,
  pp. 45--61, 2020.

\bibitem{20}
S.~Samarakoon, M.~Bennis, W.~Saad, and M.~Debbah, ``Distributed {Federated}
  {Learning} for {Ultra}-{Reliable} {Low}-{Latency} {Vehicular}
  {Communications},'' \emph{IEEE Transactions on Communications}, vol.~68,
  no.~2, pp. 1146--1159, Feb. 2020.

\bibitem{21}
N.~H. Tran, W.~Bao, A.~Zomaya, M.~N.~H. Nguyen, and C.~S. Hong, ``Federated
  {Learning} over {Wireless} {Networks}: {Optimization} {Model} {Design} and
  {Analysis},'' in \emph{Proceedings of the {IEEE} {Conference} on {Computer}
  {Communications} ({INFOCOM})}, Apr. 2019, pp. 1387--1395.

\bibitem{22}
D.~C. Nguyen, P.~N. Pathirana, M.~Ding, and A.~Seneviratne,
  ``\BIBforeignlanguage{en}{Blockchain for {5G} and {Beyond} {Networks}: {A}
  {State} of {The} {Art} {Survey}},'' \emph{\BIBforeignlanguage{en}{Journal of
  Network and Computer Applications}}, vol. 166, p. 102693, Sep. 2020.

\bibitem{508}
M.~Fang, X.~Cao, J.~Jia, and N.~Gong, ``\BIBforeignlanguage{en}{Local {Model}
  {Poisoning} {Attacks} to {Byzantine}-{Robust} {Federated} {Learning}},'' in
  \emph{\BIBforeignlanguage{en}{29th $\{$USENIX$\}$ Security Symposium
  ($\{$USENIX$\}$ Security 20)}}, 2020, pp. 1605--1622.

\bibitem{505}
V.~Mothukuri, R.~M. Parizi, S.~Pouriyeh, Y.~Huang, A.~Dehghantanha, and
  G.~Srivastava, ``\BIBforeignlanguage{en}{A survey on security and privacy of
  federated learning},'' \emph{\BIBforeignlanguage{en}{Future Generation
  Computer Systems}}, vol. 115, pp. 619--640, Feb. 2021.

\bibitem{506}
K.~Wei, J.~Li, M.~Ding, C.~Ma, H.~H. Yang, F.~Farokhi, S.~Jin, T.~Q.~S. Quek,
  and H.~V. Poor, ``Federated {Learning} {With} {Differential} {Privacy}:
  {Algorithms} and {Performance} {Analysis},'' \emph{IEEE Transactions on
  Information Forensics and Security}, vol.~15, pp. 3454--3469, Apr. 2020.

\bibitem{511}
D.~C. Nguyen, P.~N. Pathirana, M.~Ding, and A.~Seneviratne, ``Blockchain and
  {Edge} {Computing} for {Decentralized} {EMRs} {Sharing} in {Federated}
  {Healthcare},'' in \emph{{GLOBECOM} 2020 - 2020 {IEEE} {Global}
  {Communications} {Conference}}, Taipei, Taiwan, Dec. 2020, pp. 1--6.

\bibitem{25}
H.~Kim, J.~Park, M.~Bennis, and S.-L. Kim, ``Blockchained {On}-{Device}
  {Federated} {Learning},'' \emph{IEEE Communications Letters}, vol.~24, no.~6,
  pp. 1279--1283, Jun. 2020.

\bibitem{27}
S.~R. Pokhrel and J.~Choi, ``Federated {Learning} {With} {Blockchain} for
  {Autonomous} {Vehicles}: {Analysis} and {Design} {Challenges},'' \emph{IEEE
  Transactions on Communications}, vol.~68, no.~8, pp. 4734--4746, Aug. 2020.

\bibitem{29}
Y.~Lu, X.~Huang, K.~Zhang, S.~Maharjan, and Y.~Zhang, ``Low-latency {Federated}
  {Learning} and {Blockchain} for {Edge} {Association} in {Digital} {Twin}
  empowered {6G} {Networks},'' \emph{IEEE Transactions on Industrial
  Informatics}, 2020.

\bibitem{53}
Y.~Lu, X.~Huang, K.~Zhang, and Maharjan, ``Blockchain {Empowered}
  {Asynchronous} {Federated} {Learning} for {Secure} {Data} {Sharing} in
  {Internet} of {Vehicles},'' \emph{IEEE Transactions on Vehicular Technology},
  vol.~69, no.~4, pp. 4298--4311, Apr. 2020.

\bibitem{8}
F.~Sattler, S.~Wiedemann, K.-R. Müller, and W.~Samek, ``Robust and
  {Communication}-{Efficient} {Federated} {Learning} {From} {Non}-i.i.d.
  {Data},'' \emph{IEEE Transactions on Neural Networks and Learning Systems},
  pp. 1--14, 2019.

\bibitem{nguyen2019integration}
D.~C. Nguyen, P.~N. Pathirana, M.~Ding, and A.~Seneviratne, ``Integration of
  {Blockchain} and {Cloud} of {Things}: {Architecture}, {Applications} and
  {Challenges},'' \emph{IEEE Communications Surveys \& Tutorials}, vol.~22,
  no.~4, pp. 2521--2549, Aug. 2020.

\bibitem{add1}
H.~Kim, S.-H. Kim, J.~Y. Hwang, and C.~Seo, ``Efficient {Privacy}-{Preserving}
  {Machine} {Learning} for {Blockchain} {Network},'' \emph{IEEE Access},
  vol.~7, pp. 136\,481--136\,495, 2019.

\bibitem{512}
Q.~Lu and X.~Xu, ``Adaptable {Blockchain}-{Based} {Systems}: {A} {Case} {Study}
  for {Product} {Traceability},'' \emph{IEEE Software}, vol.~34, no.~6, pp.
  21--27, Nov. 2017.

\bibitem{513}
S.~Liaskos, B.~Wang, and N.~Alimohammadi, ``Blockchain {Networks} as {Adaptive}
  {Systems},'' in \emph{2019 {IEEE}/{ACM} 14th {International} {Symposium} on
  {Software} {Engineering} for {Adaptive} and {Self}-{Managing} {Systems}
  ({SEAMS})}, Montreal, QC, Canada, May 2019, pp. 139--145.

\bibitem{26}
L.~Wang, W.~Wang, and B.~Li, ``{CMFL}: {Mitigating} {Communication} {Overhead}
  for {Federated} {Learning},'' in \emph{Proceedings of the {IEEE} 39th
  {International} {Conference} on {Distributed} {Computing} {Systems}
  ({ICDCS})}, Jul. 2019, pp. 954--964.

\bibitem{28}
J.~Mills, J.~Hu, and G.~Min, ``Communication-{Efficient} {Federated} {Learning}
  for {Wireless} {Edge} {Intelligence} in {IoT},'' \emph{IEEE Internet of
  Things Journal}, vol.~7, no.~7, pp. 5986--5994, Jul. 2020.

\bibitem{30}
F.~Yang, W.~Zhou, Q.~Wu, R.~Long, N.~N. Xiong, and M.~Zhou, ``Delegated {Proof}
  of {Stake} {With} {Downgrade}: {A} {Secure} and {Efficient} {Blockchain}
  {Consensus} {Algorithm} {With} {Downgrade} {Mechanism},'' \emph{IEEE Access},
  vol.~7, pp. 118\,541--118\,555, 2019.

\bibitem{31}
Y.~Liu, K.~Wang, Y.~Lin, and W.~Xu, ``{LightChain}: A lightweight blockchain
  system for industrial {Internet} of things,'' \emph{IEEE Transactions on
  Industrial Informatics}, vol.~15, no.~6, pp. 3571--3581, Jun. 2019.

\bibitem{32}
N.~Q. Hieu, T.~T. Anh, N.~C. Luong, D.~Niyato, D.~I. Kim, and E.~Elmroth,
  ``Resource {Management} for {Blockchain}-enabled {Federated} {Learning}: {A}
  {Deep} {Reinforcement} {Learning} {Approach},'' \emph{arXiv:2004.04104}, May
  2020.

\bibitem{33}
H.~T. Nguyen, N.~C. Luong, J.~Zhao, C.~Yuen, and D.~Niyato, ``Resource
  {Allocation} in {Mobility}-{Aware} {Federated} {Learning} {Networks}: {A}
  {Deep} {Reinforcement} {Learning} {Approach},'' in \emph{IEEE 3rd World Forum
  on Internet of Things {(WF-IoT)}}, 2020.

\bibitem{34}
Y.~Lu, X.~Huang, K.~Zhang, S.~Maharjan, and Y.~Zhang,
  ``Communication-{Efficient} {Federated} {Learning} for {Digital} {Twin}
  {Edge} {Networks} in {Industrial} {IoT},'' \emph{IEEE Transactions on
  Industrial Informatics}, 2020.

\bibitem{35}
Z.~Liao and R.~Couillet, ``A {Large} {Dimensional} {Analysis} of {Least}
  {Squares} {Support} {Vector} {Machines},'' \emph{IEEE Transactions on Signal
  Processing}, vol.~67, no.~4, pp. 1065--1074, Feb. 2019.

\bibitem{36}
M.~Liu, F.~R. Yu, Y.~Teng, V.~C.~M. Leung, and M.~Song, ``Distributed
  {Resource} {Allocation} in {Blockchain}-{Based} {Video} {Streaming} {Systems}
  {With} {Mobile} {Edge} {Computing},'' \emph{IEEE Transactions on Wireless
  Communications}, vol.~18, no.~1, pp. 695--708, Jan. 2019.

\bibitem{37}
Y.~Zhan, P.~Li, Z.~Qu, D.~Zeng, and S.~Guo, ``A {Learning}-{Based} {Incentive}
  {Mechanism} for {Federated} {Learning},'' \emph{IEEE Internet of Things
  Journal}, vol.~7, no.~7, pp. 6360--6368, Jul. 2020.

\bibitem{38}
K.~Toyoda and A.~N. Zhang, ``Mechanism {Design} for {An} {Incentive}-aware
  {Blockchain}-enabled {Federated} {Learning} {Platform},'' in
  \emph{Proceedings of the {IEEE} {International} {Conference} on {Big} {Data}
  ({Big} {Data})}, Dec. 2019, pp. 395--403.

\bibitem{39}
M.~H. Ur~Rehman, K.~Salah, E.~Damiani, and D.~Svetinovic, ``Towards
  {Blockchain}-{Based} {Reputation}-{Aware} {Federated} {Learning},'' in
  \emph{Proceedings of the {IEEE} {Conference} on {Computer} {Communications}
  {Workshops} ({INFOCOM} {WKSHPS})}, Jul. 2020, pp. 183--188.

\bibitem{40}
N.~Baranwal~Somy, K.~Kannan, V.~Arya, S.~Hans, A.~Singh, P.~Lohia, and
  S.~Mehta, ``Ownership {Preserving} {AI} {Market} {Places} {Using}
  {Blockchain},'' in \emph{Proceedings of the 2019 {IEEE} {International}
  {Conference} on {Blockchain} ({Blockchain})}, Jul. 2019, pp. 156--165.

\bibitem{41}
X.~Bao, C.~Su, Y.~Xiong, W.~Huang, and Y.~Hu, ``{FLChain}: {A} {Blockchain} for
  {Auditable} {Federated} {Learning} with {Trust} and {Incentive},'' in
  \emph{Proceedings of the 5th {International} {Conference} on {Big} {Data}
  {Computing} and {Communications} ({BIGCOM})}, Aug. 2019, pp. 151--159.

\bibitem{42}
J.~Weng, J.~Weng, J.~Zhang, M.~Li, Y.~Zhang, and W.~Luo, ``{DeepChain}:
  {Auditable} and {Privacy}-{Preserving} {Deep} {Learning} with
  {Blockchain}-based {Incentive},'' \emph{IEEE Transactions on Dependable and
  Secure Computing}, 2019.

\bibitem{43}
I.~Martinez, S.~Francis, and A.~S. Hafid, ``Record and {Reward} {Federated}
  {Learning} {Contributions} with {Blockchain},'' in \emph{Proceedings of the
  {International} {Conference} on {Cyber}-{Enabled} {Distributed} {Computing}
  and {Knowledge} {Discovery} ({CyberC})}, Oct. 2019, pp. 50--57.

\bibitem{517}
D.~C. Nguyen, P.~N. Pathirana, M.~Ding, and A.~Seneviratne, ``Blockchain for
  {Secure} {EHRs} {Sharing} of {Mobile} {Cloud} {Based} {E}-{Health}
  {Systems},'' \emph{IEEE Access}, vol.~7, pp. 66\,792--66\,806, 2019.

\bibitem{44}
Y.~Jiao, P.~Wang, D.~Niyato, and K.~Suankaewmanee, ``Auction {Mechanisms} in
  {Cloud}/{Fog} {Computing} {Resource} {Allocation} for {Public} {Blockchain}
  {Networks},'' \emph{IEEE Transactions on Parallel and Distributed Systems},
  vol.~30, no.~9, pp. 1975--1989, Sep. 2019.

\bibitem{49}
R.~Schmid, B.~Pfitzner, J.~Beilharz, B.~Arnrich, and A.~Polze, ``Tangle
  {Ledger} for {Decentralized} {Learning},'' in \emph{{IEEE} {International}
  {Parallel} and {Distributed} {Processing} {Symposium} {Workshops}
  ({IPDPSW})}, May 2020, pp. 852--859.

\bibitem{45}
Y.~Qu, L.~Gao, T.~H. Luan, Y.~Xiang, S.~Yu, B.~Li, and G.~Zheng,
  ``Decentralized {Privacy} {Using} {Blockchain}-{Enabled} {Federated}
  {Learning} in {Fog} {Computing},'' \emph{IEEE Internet of Things Journal},
  vol.~7, no.~6, pp. 5171--5183, Jun. 2020.

\bibitem{48}
Q.~Wang, Y.~Guo, X.~Wang, T.~Ji, L.~Yu, and P.~Li, ``{AI} at the {Edge}:
  {Blockchain}-{Empowered} {Secure} {Multiparty} {Learning} with
  {Heterogeneous} {Models},'' \emph{IEEE Internet of Things Journal}, 2020.

\bibitem{46}
P.~C.~M. Arachchige, P.~Bertok, I.~Khalil, D.~Liu, S.~Camtepe, and
  M.~Atiquzzaman, ``A {Trustworthy} {Privacy} {Preserving} {Framework} for
  {Machine} {Learning} in {Industrial} {IoT} {Systems},'' \emph{IEEE
  Transactions on Industrial Informatics}, vol.~16, no.~9, pp. 6092--6102, Sep.
  2020.

\bibitem{515}
Y.~Zhao, J.~Zhao, M.~Yang, T.~Wang, N.~Wang, L.~Lyu, D.~Niyato, and K.-Y. Lam,
  ``Local {Differential} {Privacy} based {Federated} {Learning} for {Internet}
  of {Things},'' \emph{IEEE Internet of Things Journal}, 2020.

\bibitem{516}
S.~Ma, Y.~Cao, and L.~Xiong, ``Transparent {Contribution} {Evaluation} for
  {Secure} {Federated} {Learning} on {Blockchain},'' Jan. 2021, arXiv:
  2101.10572.

\bibitem{47}
S.~Lugan, P.~Desbordes, E.~Brion, L.~X. Ramos~Tormo, A.~Legay, and B.~Macq,
  ``Secure {Architectures} {Implementing} {Trusted} {Coalitions} for
  {Blockchained} {Distributed} {Learning} ({TCLearn}),'' \emph{IEEE Access},
  vol.~7, pp. 181\,789--181\,799, 2019.

\bibitem{514}
X.~Li, Y.~Wang, J.~Song, Y.~Liu, X.~Zhang, K.~Zhou, and C.~Li,
  ``\BIBforeignlanguage{en}{A low cost and un-cancelled laplace noise based
  differential privacy algorithm for spatial decompositions},''
  \emph{\BIBforeignlanguage{en}{World Wide Web}}, vol.~23, no.~1, pp. 549--572,
  Jan. 2020.

\bibitem{50}
S.~Zhang and J.-H. Lee, ``Double-{Spending} {With} a {Sybil} {Attack} in the
  {Bitcoin} {Decentralized} {Network},'' \emph{IEEE Transactions on Industrial
  Informatics}, vol.~15, no.~10, pp. 5715--5722, Oct. 2019.

\bibitem{add2}
L.~T. Phong, Y.~Aono, T.~Hayashi, L.~Wang, and S.~Moriai,
  ``Privacy-{Preserving} {Deep} {Learning} via {Additively} {Homomorphic}
  {Encryption},'' \emph{IEEE Transactions on Information Forensics and
  Security}, vol.~13, no.~5, pp. 1333--1345, May 2018.

\bibitem{add3}
Y.~Chen, X.~Qin, J.~Wang, C.~Yu, and W.~Gao, ``{FedHealth}: {A} {Federated}
  {Transfer} {Learning} {Framework} for {Wearable} {Healthcare},'' \emph{IEEE
  Intelligent Systems}, vol.~35, no.~4, pp. 83--93, Jul. 2020.

\bibitem{61}
Y.~Wang, Z.~Su, N.~Zhang, and A.~Benslimane, ``Learning in the {Air}: {Secure}
  {Federated} {Learning} for {UAV}-{Assisted} {Crowdsensing},'' \emph{IEEE
  Transactions on Network Science and Engineering}, 2020.

\bibitem{51}
Q.-V. Pham, F.~Fang, V.~N. Ha, M.~J. Piran, M.~Le, L.~B. Le, W.-J. Hwang, and
  Z.~Ding, ``A {Survey} of {Multi}-{Access} {Edge} {Computing} in {5G} and
  {Beyond}: {Fundamentals}, {Technology} {Integration}, and
  {State}-of-the-{Art},'' \emph{IEEE Access}, vol.~8, pp. 116\,974--117\,017,
  2020.

\bibitem{52}
Y.~Lu, X.~Huang, Y.~Dai, S.~Maharjan, and Y.~Zhang, ``Blockchain and
  {Federated} {Learning} for {Privacy}-{Preserved} {Data} {Sharing} in
  {Industrial} {IoT},'' \emph{IEEE Transactions on Industrial Informatics},
  vol.~16, no.~6, pp. 4177--4186, Jun. 2020.

\bibitem{54}
H.~Chai, S.~Leng, Y.~Chen, and K.~Zhang, ``A {Hierarchical}
  {Blockchain}-{Enabled} {Federated} {Learning} {Algorithm} for {Knowledge}
  {Sharing} in {Internet} of {Vehicles},'' \emph{IEEE Transactions on
  Intelligent Transportation Systems}, 2020.

\bibitem{55}
B.~Yin, H.~Yin, Y.~Wu, and Z.~Jiang, ``{FDC}: {A} {Secure} {Federated} {Deep}
  {Learning} {Mechanism} for {Data} {Collaborations} in the {Internet} of
  {Things},'' \emph{IEEE Internet of Things Journal}, vol.~7, no.~7, pp.
  6348--6359, Jul. 2020.

\bibitem{56}
Z.~Yu, J.~Hu, G.~Min, H.~Lu, Z.~Zhao, H.~Wang, and N.~Georgalas, ``Federated
  {Learning} {Based} {Proactive} {Content} {Caching} in {Edge} {Computing},''
  in \emph{{IEEE} {Global} {Communications} {Conference} ({GLOBECOM})}, Dec.
  2018, pp. 1--6.

\bibitem{57}
L.~Cui, X.~Su, Z.~Ming, Z.~Chen, S.~Yang, Y.~Zhou, and W.~Xiao, ``{CREAT}:
  {Blockchain}-assisted {Compression} {Algorithm} of {Federated} {Learning} for
  {Content} {Caching} in {Edge} {Computing},'' \emph{IEEE Internet of Things
  Journal}, 2020.

\bibitem{58}
N.~Garg, M.~Sellathurai, V.~Bhatia, B.~N. Bharath, and T.~Ratnarajah, ``Online
  {Content} {Popularity} {Prediction} and {Learning} in {Wireless} {Edge}
  {Caching},'' \emph{IEEE Transactions on Communications}, vol.~68, no.~2, pp.
  1087--1100, Feb. 2020.

\bibitem{59}
X.~Wang, Y.~Han, C.~Wang, Q.~Zhao, X.~Chen, and M.~Chen, ``In-{Edge} {AI}:
  {Intelligentizing} {Mobile} {Edge} {Computing}, {Caching} and {Communication}
  by {Federated} {Learning},'' \emph{IEEE Network}, vol.~33, no.~5, pp.
  156--165, Sep. 2019.

\bibitem{60}
Z.~Zhou, H.~Liao, B.~Gu, K.~M.~S. Huq, S.~Mumtaz, and J.~Rodriguez, ``Robust
  {Mobile} {Crowd} {Sensing}: {When} {Deep} {Learning} {Meets} {Edge}
  {Computing},'' \emph{IEEE Network}, vol.~32, no.~4, pp. 54--60, Jul. 2018.

\bibitem{62}
V.~Kulkarni, M.~Kulkarni, and A.~Pant, ``Survey of {Personalization}
  {Techniques} for {Federated} {Learning},'' \emph{arXiv:2003.08673}, Mar.
  2020.

\bibitem{63}
S.~R. Pandey, N.~H. Tran, M.~Bennis, Y.~K. Tun, A.~Manzoor, and C.~S. Hong, ``A
  {Crowdsourcing} {Framework} for {On}-{Device} {Federated} {Learning},''
  \emph{IEEE Transactions on Wireless Communications}, vol.~19, no.~5, pp.
  3241--3256, May 2020.

\bibitem{64}
Y.~Zhao, J.~Zhao, L.~Jiang, R.~Tan, D.~Niyato, Z.~Li, L.~Lyu, and Y.~Liu,
  ``Privacy-{Preserving} {Blockchain}-{Based} {Federated} {Learning} for {IoT}
  {Devices},'' \emph{IEEE Internet of Things Journal}, pp. 1--1, Aug. 2020.

\bibitem{65}
X.~Li, P.~Jiang, T.~Chen, X.~Luo, and Q.~Wen, ``\BIBforeignlanguage{en}{A
  survey on the security of blockchain systems},''
  \emph{\BIBforeignlanguage{en}{Future Generation Computer Systems}}, vol. 107,
  pp. 841--853, Jun. 2020.

\bibitem{challenge7}
S.~Wang, C.~Wang, and Q.~Hu, ``Corking by {Forking}: {Vulnerability} {Analysis}
  of {Blockchain},'' in \emph{Proceedings of the {IEEE} {Conference} on
  {Computer} {Communications}}, Paris, France, Apr. 2019, pp. 829--837.

\bibitem{66}
S.~Zhang and J.-H. Lee, ``Mitigations on {Sybil}-based {Double}-spend {Attacks}
  in {Bitcoin},'' \emph{IEEE Consumer Electronics Magazine}, 2020.

\bibitem{zhanggan2020}
J.~Zhang, J.~Zhang, J.~Chen, and S.~Yu, ``{GAN} {Enhanced} {Membership}
  {Inference}: {A} {Passive} {Local} {Attack} in {Federated} {Learning},'' in
  \emph{Proceedings of the 2020 {IEEE} {International} {Conference} on
  {Communications} ({ICC})}, Dublin, Ireland, Jun. 2020, pp. 1--6.

\bibitem{67}
P.~Silva, ``Impact of {Geo}-{Distribution} and {Mining} {Pools} on
  {Blockchains}: {A} {Study} of {Ethereum} - {Practical} {Experience} {Report}
  and {Ongoing} {PhD} {Work},'' in \emph{Proceedings of the 50th {Annual}
  {IEEE}-{IFIP} {International} {Conference} on {Dependable} {Systems} and
  {Networks}-{Supplemental} {Volume} ({DSN}-{S})}, Jun. 2020, pp. 73--74.

\bibitem{500}
N.~Shlezinger, M.~Chen, Y.~C. Eldar, H.~V. Poor, and S.~Cui, ``{UVeQFed}:
  {Universal} {Vector} {Quantization} for {Federated} {Learning},'' \emph{IEEE
  Transactions on Signal Processing}, vol.~69, pp. 500--514, Dec. 2021.

\bibitem{501}
D.~Jhunjhunwala, A.~Gadhikar, G.~Joshi, and Y.~C. Eldar, ``Adaptive
  {Quantization} of {Model} {Updates} for {Communication}-{Efficient}
  {Federated} {Learning},'' Feb. 2021, arXiv: 2102.04487.

\bibitem{502}
Z.~Chen, P.~Tian, W.~Liao, and W.~Yu, ``Zero {Knowledge} {Clustering} {Based}
  {Adversarial} {Mitigation} in {Heterogeneous} {Federated} {Learning},''
  \emph{IEEE Transactions on Network Science and Engineering}, 2020.

\bibitem{503}
D.~Rothchild, A.~Panda, E.~Ullah, N.~Ivkin, I.~Stoica, V.~Braverman,
  J.~Gonzalez, and R.~Arora, ``\BIBforeignlanguage{en}{{FetchSGD}:
  {Communication}-{Efficient} {Federated} {Learning} with {Sketching}},'' in
  \emph{\BIBforeignlanguage{en}{International {Conference} on {Machine}
  {Learning}}}, Nov. 2020, pp. 8253--8265.

\bibitem{504}
A.~Ghosh, J.~Hong, D.~Yin, and K.~Ramchandran, ``Robust {Federated} {Learning}
  in a {Heterogeneous} {Environment},'' Oct. 2019, arXiv: 1906.06629.

\bibitem{challenge1}
R.~Qin, Y.~Yuan, S.~Wang, and F.-Y. Wang, ``Economic {Issues} in {Bitcoin}
  {Mining} and {Blockchain} {Research},'' in \emph{2018 {IEEE} {Intelligent}
  {Vehicles} {Symposium} ({IV})}, Changshu, Jun. 2018, pp. 268--273.

\bibitem{challenge2}
Y.~Wang, Z.~Su, and N.~Zhang, ``{BSIS}: {Blockchain}-{Based} {Secure}
  {Incentive} {Scheme} for {Energy} {Delivery} in {Vehicular} {Energy}
  {Network},'' \emph{IEEE Transactions on Industrial Informatics}, vol.~15,
  no.~6, pp. 3620--3631, Jun. 2019.

\bibitem{70}
T.~Nishio and R.~Yonetani, ``Client {Selection} for {Federated} {Learning} with
  {Heterogeneous} {Resources} in {Mobile} {Edge},'' in \emph{Proceedings of the
  {IEEE} {International} {Conference} on {Communications} ({ICC})}, May 2019,
  pp. 1--7.

\bibitem{challenge3}
C.~Tang, L.~Wu, G.~Wen, and Z.~Zheng, ``Incentivizing {Honest} {Mining} in
  {Blockchain} {Networks}: {A} {Reputation} {Approach},'' \emph{IEEE
  Transactions on Circuits and Systems II: Express Briefs}, vol.~67, no.~1, pp.
  117--121, Jan. 2020.

\bibitem{challenge4}
E.~K. Wang, Z.~Liang, C.-M. Chen, S.~Kumari, and M.~K. Khan,
  ``\BIBforeignlanguage{en}{{PoRX}: {A} {Reputation} {Incentive} {Scheme} for
  {Blockchain} {Consensus} of {IIoT}},'' \emph{\BIBforeignlanguage{en}{Future
  Generation Computer Systems}}, vol. 102, pp. 140--151, Jan. 2020.

\bibitem{challenge5}
M.~Chuan, L.~Jun, D.~Ming, S.~Long, W.~Taotao, H.~Zhu, and P.~H.~Vincent,
  ``{When} {Federated} {Learning} {Meets} {Blockchain}: {A} {New} {Distributed}
  {Learning} {Paradigm},'' \emph{arXiv: 2009.09338}, Sep. 2020.

\bibitem{challenge6}
A.~Diro, H.~Reda, N.~Chilamkurti, A.~Mahmood, N.~Zaman, and Y.~Nam,
  ``Lightweight {Authenticated}-{Encryption} {Scheme} for {Internet} of
  {Things} {Based} on {Publish}-{Subscribe} {Communication},'' \emph{IEEE
  Access}, vol.~8, pp. 60\,539--60\,551, 2020.

\bibitem{68}
W.~Shi, S.~Zhou, and Z.~Niu, ``Device {Scheduling} with {Fast} {Convergence}
  for {Wireless} {Federated} {Learning},'' in \emph{Proceedings of the {IEEE}
  {International} {Conference} on {Communications} ({ICC})}, Jun. 2020, pp.
  1--6.

\bibitem{69}
S.~Biswas, K.~Sharif, F.~Li, S.~Maharjan, S.~P. Mohanty, and Y.~Wang, ``{PoBT}:
  {A} {Lightweight} {Consensus} {Algorithm} for {Scalable} {IoT} {Business}
  {Blockchain},'' \emph{IEEE Internet of Things Journal}, vol.~7, no.~3, pp.
  2343--2355, Mar. 2020.

\bibitem{509}
D.~C. Nguyen, P.~N. Pathirana, M.~Ding, and A.~Seneviratne,
  ``Privacy-{Preserved} {Task} {Offloading} in {Mobile} {Blockchain} {With}
  {Deep} {Reinforcement} {Learning},'' \emph{IEEE Transactions on Network and
  Service Management}, vol.~17, no.~4, pp. 2536--2549, Dec. 2020.

\bibitem{510}
H.~Yang, Z.~Xiong, J.~Zhao, T.~D. Niyato, C.~Yuen, and R.~Deng, ``Deep
  {Reinforcement} {Learning} {Based} {Massive} {Access} {Management} for
  {Ultra}-{Reliable} {Low}-{Latency} {Communications},'' \emph{IEEE
  Transactions on Wireless Communications}, 2021.

\end{thebibliography}
